\documentclass{emulateapj}

\def\kms{\hbox{$~$km$~$s$^{-1}$}}
\def\l{\ifmmode\lambda\else$\lambda$\fi}
\def\snid{\ifmmode{\rm SNID}\else{SNID}\fi}

\def\catwo{Ca~{\sc ii}}

\def\fetwo{Fe~{\sc ii}}
\def\fethree{Fe~{\sc iii}}

\def\heone{He~{\sc i}}

\def\stwo{S~{\sc ii}}
\def\sitwo{Si~{\sc ii}}

\slugcomment{The Astrophysical Journal, 666:1024--1047, 2007 September 10}
\shorttitle{Type, Redshift, \& Age of SN spectra}
\shortauthors{Blondin \& Tonry}

\begin{document}

\title{Determining the Type, Redshift, and Age of a Supernova Spectrum}

\author{St{\'e}phane Blondin}
\affil{Harvard-Smithsonian Center for Astrophysics, Cambridge, MA 02138}
\email{sblondin@cfa.harvard.edu}

\and

\author{John L. Tonry}
\affil{Institute for Astronomy, University of Hawaii, Honolulu, HI 96822}
\email{jt@ifa.hawaii.edu}


\begin{abstract}
We present an algorithm to identify the type of an SN spectrum
and to determine its redshift and age. This algorithm, based on the 
correlation techniques of Tonry \& Davis, is implemented in the
Supernova Identification (\snid) code. It is used by members of 
ongoing high-redshift SN searches to distinguish between
type Ia and type Ib/c SNe, and to identify ``peculiar'' SNe~Ia. We
develop a diagnostic to quantify the quality of a correlation between
the input and template spectra, which enables a formal evaluation of
the associated redshift error. Furthermore, by comparing the
correlation redshifts obtained using \snid\ with those determined from
narrow lines in the SN host galaxy spectrum, we show that
accurate redshifts (with a typical error $\sigma_z \lesssim 0.01$) can
be determined for SNe~Ia without a spectrum of the host galaxy. 
Last, the age of an input spectrum is determined with a
typical accuracy $\sigma_t \lesssim 3$ days, shown here by using
high-redshift SNe~Ia with well-sampled light curves. The success of
the correlation technique 
confirms the similarity of some SNe Ia at low and high redshifts.
The \snid\ code,
which will be made available to the community, can also be used for 
comparative studies of SN spectra, as well as comparisons
between data and models.
\end{abstract}



\section{Introduction}\label{Sect:intro}

Supernovae (SNe) play a major role in the recent revival of observational
cosmology. It is through comparison of high-redshift Type Ia
supernova (SN~Ia) magnitudes with those at low-redshift
\citep{Hamuy/etal:1996,Riess/etal:1999,Jha/etal:2006a} that two teams
independently found the present rate of the universal expansion to be
accelerating \citep{Riess/etal:1998a,Perlmutter/etal:1999}. This
astonishing result has been confirmed in subsequent years out to
redshift $z \lesssim 1$
\citep{Tonry/etal:2003,Knop/etal:2003,Barris/etal:2004}, but also at
higher redshifts where the universal expansion is in a decelerating
phase \citep{Riess/etal:2004}. Currently, two ongoing projects have
the more ambitious goal to measure the equation-of-state parameter,
$w$, of the ``dark energy'' that drives the expansion: the ESSENCE
(Equation of State: SupErNovae trace Cosmic Expansion;
\citealt{Miknaitis/etal:2007,Wood-Vasey/etal:2007}) and SNLS
(SuperNova Legacy Survey; \citealt{Astier/etal:2006}) projects. Both
teams have published their initial results, which indicate that $w$,
if constant, is consistent with a cosmological constant ($w=-1$).

The success of these cosmological experiments depends, among other
things, on the assurance that the supernovae in the sample are of the
correct type, namely, SNe~Ia. The classification of supernovae is
based on their optical spectra around maximum light (for a review see
\citealt{Filippenko:1997}). At high redshifts, obtaining
sufficiently high signal-to-noise ratio (S/N) spectra of such objects
requires 1--2\,hr integrations at 6.5--10\,m-class telescopes (see,
e.g., \citealt{Matheson/etal:2005}), and constitutes the limiting
factor for these experiments. Recently, alternative classification
methods based on photometry alone have been suggested
\citep{Poznanski/Maoz/Gal-Yam:2006,Kuznetsova/Connolly:2007,Kunz/etal:2006},
in anticipation of the next generation of wide-field all-sky surveys
optimized for the detection of transient events
(Dark Energy Survey, \citealt{Frieman:2004}; Pan-STARRS,
\citealt{Kaiser/etal:2002}; SKYMAPPER, \citealt{Schmidt/etal:2005};
ALPACA, \citealt{Crotts/etal:2006}; LSST,
\citealt{Tyson/Angel:2001}). Inclusion of supernovae that are 
of a different type leads to biased cosmological parameters
\citep{Homeier:2005}. Exclusion of genuine SNe~Ia from the sample
leads to increased statistical errors on these same parameters.

The secure classification of supernovae is a challenge at all
redshifts, however. Even with high S/N spectra, the distinction
between supernovae of different types (or between subtypes within a
given type) can pose problems. This points to the inadequacy of the
present purely empirical SN classification scheme in establishing
distinct classes of supernovae, whose observational properties can be
directly traced back to an explosion mechanism and a progenitor
system. The two major types of supernovae are defined based on the
presence (Type II) or absence (Type I) of hydrogen in their
spectra, a distinction that does not reflect the differences in
their explosion mechanisms and progenitors: through the
thermonuclear disruption of a carbon-oxygen white dwarf star (Type
Ia), or through the collapse of the degenerate core of a massive
star (Types Ib, Ic, and II). For the latter case, it is now thought
that a there exists a continuity of events between the types
II$\rightarrow$Ib$\rightarrow$Ic, corresponding to increasing mass
loss of the outer envelope of the progenitor star prior to explosion
\citep{Chevalier:2006}. SNe~IIb are an
intermediate case between Type II and Type Ib, and illustrate the
tendency of some supernovae to ``evolve'' from one type to
another. SNe~Ic supernovae are only defined by the {\it absence} of
elements 
(hydrogen and helium; although see \citealt{Elmhamdi/etal:2006} for
the presence of hydrogen in SNe~Ib/c) in their atmospheres and thus
form a heterogeneous class--- which includes the supernovae associated
with gamma-ray bursts
\citep{Kulkarni/etal:1998,Matheson/etal:2003}. The classification 
scheme is further complicated by ``peculiar'' sub-classes of events
associated with the four types (Ia, Ib, Ic, and
II). Nonetheless, this classification scheme provides a means to keep
track of general spectroscopic properties associated with the many
supernovae discovered each year 
(more than 550 in 2006 according
to the International Astronomical Union\footnote{{\tt
    http://www.cfa.harvard.edu/iau/lists/Supernovae.html}}) and is
useful for comparative studies of supernovae with similar
characteristics.

The spectrum of a supernova also contains information on its redshift
and age (
defined as the number of days from
maximum light in a given filter). Knowledge of the SN redshift is
necessary for the use of SNe~Ia as distance indicators (although see
\citealt{Barris/Tonry:2004} for redshift-independent distances), and
is usually determined using narrow lines in the spectrum of the host
galaxy. When such a spectrum is unavailable, one has to rely on
comparison with SN template spectra for determining the redshift,
although we note that \citet{Wang:2007} has recently presented a 
purely photometric redshift estimator for SNe~Ia, albeit with 3--5
times larger errors.
The SN age is usually determined (to within 1 day) using a
well-sampled light curve, but a single spectrum can also provide a
good estimate (to within 2--3 days for SNe~Ia;
\citealt{Riess/etal:1997}), since the relative strengths and
wavelength location of spectral features evolve 
significantly on the timescale of days. Knowledge of
the age of the supernova and its apparent magnitude and color at a
single epoch can yield a distance measurement accurate to $\sim 10\%$  
\citep{Riess/etal:1998b}. Moreover, comparison of spectral and
light-curve ages of high-redshift supernovae can be used to test the
expected time-dilation factor of $(1+z)$, where $z$ is the redshift,
in an expanding universe \citep{Riess/etal:1997,Foley/etal:2005}.

We have developed a tool (Supernova Identification [\snid]) to determine
the type, redshift, and age of a supernova, using a single
spectrum. The algorithm is based on the correlation techniques of 
\citet{Tonry/Davis:1979}, and relies on the comparison of an input
spectrum with a database of high-S/N template spectra. Fundamental to
the success of the correlation technique is its application to objects
that have counterparts in the template database. We briefly
describe the cross-correlation technique in the next section, before
presenting the algorithm for determining the redshift
(\S~\ref{Sect:z}). We then briefly comment on the composition of our
spectral database (\S~\ref{Sect:sniddb}), before testing the accuracy
of correlation redshifts and ages using \snid\ in
\S~\ref{Sect:zt}. Last, in \S~\ref{Sect:type}, we tackle the issue of
supernova classification by focusing on specific examples, some of
which are particularly relevant to SN searches at high redshifts.


\section{Cross-correlation Formalism}\label{Sect:xcorform}

The cross-correlation method presented in this section is extensively
discussed by \citet{Tonry/Davis:1979}, where it is exclusively applied
to galaxy spectra. We reproduce part of this discussion here to
highlight the specificity of determining supernova (as opposed to
galaxy) redshifts.

Sections \ref{Sect:defs} and \ref{Sect:xcorz} are rather
technical, while \S~\ref{Sect:preproc} presents the more practical
aspects of spectrum pre-processing necessary for the
cross-correlation method.

\subsection{A Few Definitions\label{Sect:defs}}

The correlation technique is straightforward: a
supernova spectrum $s(n)$ whose redshift $z_s$ is to be found is
cross-correlated with a template spectrum (of known type and age)
$t(n)$ at zero redshift. We want to determine the $(1+z_s^\prime)$
wavelength scaling that maximizes the cross-correlation $c(n) = s(n)
\star t(n)$, where $\star$ denotes the cross-correlation product. In
practice, it is convenient to bin the spectra on a
logarithmic wavelength axis. Multiplying
the wavelength axis of $t(n)$ by a factor $(1+z)$ is equivalent
to adding a $\ln (1+z)$ shift to the logarithmic wavelength axis of
$t(n)$, i.e. a (velocity) redshift corresponds to a uniform linear
shift. Supposing we bin $s(n)$ and $t(n)$ into $N$ bins in the range
$[\l_0,\l_1]$, each wavelength coordinate $\l_{{\rm ln},n}$ is given
by

\begin{equation}
\label{Eqn:dllog}
\l_{{\rm ln},n} = \l_0\ e^{n \times d\l_{\rm ln}},
\end{equation}

\noindent
where $d\l_{\rm ln} = \ln{(\l_1/\l_0)}/N$ is the size of a
logarithmic wavelength bin, and assuming $n$ runs from 0 to $N$. We
then have:

\begin{equation}
\label{Eqn:nlnl}
n = A \ln{\l_{{\rm ln},n}} + B,
\end{equation}

\noindent
where $A = N / \ln{(\l_1/\l_0)}$ and $B = -N\ln{\l_0} /
\ln{(\l_1/\l_0)}$. In what follows we assume that $s(n)$ and $t(n)$
have been normalized such that their mean is zero (see
\S~\ref{Sect:preproc}).

For computational ease and for pre-processing purposes, the
cross-correlation is computed in Fourier space. Let $S(k)$ and $T(k)$
be the discrete Fourier transforms of the supernova and template
spectra, respectively ($k$ is the wavenumber):

\begin{equation}
S,T(k) = \sum_{n=0}^{N-1} s,t(n)\ e^{-2\pi ink/N}.
\end{equation}

\noindent
Let $\sigma_s$ and $\sigma_t$ be the rms of the
supernova and template spectrum, respectively:

\begin{equation}
\label{Eqn:sigma2}
\sigma_{s,t}^2 = \frac{1}{N} \sum_{n=0}^{N-1} s,t(n)^2.
\end{equation}

\noindent
The {\it normalized} correlation function $c(n)$ is defined as 

\begin{equation}
\label{Eqn:cnorm}
c(n) = s(n) \star t(n) = \frac{1}{N\sigma_s\sigma_t}
\sum_{m=0}^{N-1} s(m)t(m-n), 
\end{equation}

\noindent
such that if the supernova spectrum is the same as the template
spectrum, but shifted by $\delta$ logarithmic wavelength bins--- i.e. $s(n) =
t(n-\delta)$, then $c(\delta) = 1$. The Fourier transform of the
correlation function is

\begin{equation}
\label{Eqn:C(k)}
C(k) = \frac{1}{N\sigma_s\sigma_t} S(k) \overline{T(k)},
\end{equation}

\noindent
where $\overline{T(k)}$ denotes the complex conjugate of $T(k)$.

\subsection{Cross-correlation Redshifts\label{Sect:xcorz}}

Following \citet{Tonry/Davis:1979} we assume that $s(n)$ is some
multiple $\alpha$ of $t(n)$, but shifted by $\delta$ logarithmic
wavelength bins:

\begin{equation}
s(n) = \alpha t(n-\delta).
\end{equation}

\noindent
Unlike \citet{Tonry/Davis:1979}, however, we do not need to assume
that $t(n-\delta)$ is further convolved with a broadening symmetric
function [$b(n)$ in \citealt{Tonry/Davis:1979}, their eq.~6] that
accounts for galaxy stellar velocity dispersions and spectrograph
resolutions. While there exists a velocity dispersion residual in
$c(n)$ primarily due to differences in the dynamics of the expanding
envelope for different supernovae, this residual carries important
information on the age of the supernova, which we also want to
determine (see \S~\ref{Sect:zt}), and on the specificity of $s(n)$,
which is important for more general comparative supernova
studies. Second, the nominal Doppler width of a supernova spectral
``feature'' is $\sim$1--2 orders of magnitude greater than the
resolution of a typical low-resolution spectrograph.

To estimate $\alpha$ and $\delta$, we need to minimize the following
expression:

\begin{eqnarray}
\label{Eqn:chi2}
\chi^2(\alpha,\delta) &=& \sum_{n=0}^{N-1} [\alpha t(n-\delta)-s(n)]^2 \\
\label{Eqn:chi2exp}
\Rightarrow \chi^2(\alpha,\delta) &=& \alpha^2 N \sigma_t^2 -
2\alpha N \sigma_s \sigma_t c(\delta) + N \sigma_s^2,
\end{eqnarray}

\noindent
using eqs.~\ref{Eqn:sigma2} and \ref{Eqn:cnorm}. We then obtain the
condition for minimizing $\chi^2$ with respect to $\alpha$:

\begin{equation}
\frac{\partial \chi^2}{\partial \alpha} = 
2N[\alpha \sigma_t^2 - \sigma_s \sigma_t c(\delta)] = 0,
\end{equation}

\noindent
from which we derive $\alpha_{\rm min}$ satisfying the above:

\begin{equation}
\alpha_{\rm min} = \frac{\sigma_s}{\sigma_t} c(\delta).
\end{equation}

\noindent
Substituting this value for $\alpha$ back into eq.~\ref{Eqn:chi2exp},
we obtain a new expression for $\chi^2$:

\begin{equation}
\chi^2(\alpha_{\rm min},\delta) = N \sigma_s^2 [1 - c(\delta)^2].
\end{equation}

\noindent
As expected, minimizing $\chi^2$ is equivalent to maximizing the
normalized correlation function $c(\delta)$.

Thus, the input supernova spectrum $s(n)$ is cross-correlated with a
template spectrum $t(n)$, and a smooth function (here a fourth-order
polynomial, as in \citealt{Tonry/Davis:1979}) is fitted to the highest
peak in $c(n)$, whose height and center determine $\alpha$ and
$\delta$, respectively. The cross-correlation redshift is then
trivially computed as

\begin{equation}
z_s^\prime = e^{\delta \times d\l_{\rm ln}} - 1,
\end{equation}

\noindent
where $d\l_{\rm ln}$ is the logarithmic wavelength bin defined in
eq.~\ref{Eqn:dllog}. The width of the peak is a measure of the error
in $z_s^\prime$ and is of the order of the typical width of a
supernova spectral feature, modulated by the signal-to-noise ratio of
the input spectrum (see \S~\ref{Sect:zerr}).

It is important to note that we assume the noise per pixel to be
constant in the input spectrum. This is clearly not the case for
ground-based optical spectra, where sharp emission features from the
sky background leads to increased noise at specific
wavelengths. Recently, \citet{Saunders/Cannon/Sutherland:2004} found
that scaling the input spectrum by the inverse-variance yielded a
dramatic improvement in the derived cross-correlation redshifts;
specifically, \citet{Saunders/Cannon/Sutherland:2004} rewrite
eq.~\ref{Eqn:chi2} as

\begin{equation}
\chi^2(\alpha,\delta) = \sum_{n=0}^{N-1} \left[ \frac{\alpha
    t(n-\delta)-s(n)}{\sigma(n)} \right]^2,
\end{equation}

\noindent
where $\sigma(n)$ is the noise per pixel of $s(n)$, and find that this
is equivalent to simply scaling $s(n)$ by $1/\sigma(n)^2$.

This modification is well suited for determining galaxy redshifts,
since sharp features in the variance spectrum (due to sky noise) have
widths comparable to galaxy lines and hence will affect the Fourier
transform of the correlation function, $C(k)$, at similar
wavenumbers. However, we have found no such improvement for
determining supernova redshifts. This is expected since supernova
spectra consist of overlapping Doppler-broadened lines whose widths
($\sim 10000$\kms) are 1--2 orders of magnitude greater than sky
emission features. However, noise from an underlying galaxy continuum
can yield power at similar wavenumbers as $C(k)$ and can
significantly degrade the redshift accuracy when the fraction of
galaxy light in the supernova spectrum is high (see
\S~\ref{Sect:zt}).

\subsection{Pre-processing the Supernova Spectrum\label{Sect:preproc}}

\begin{figure*}
\plotone{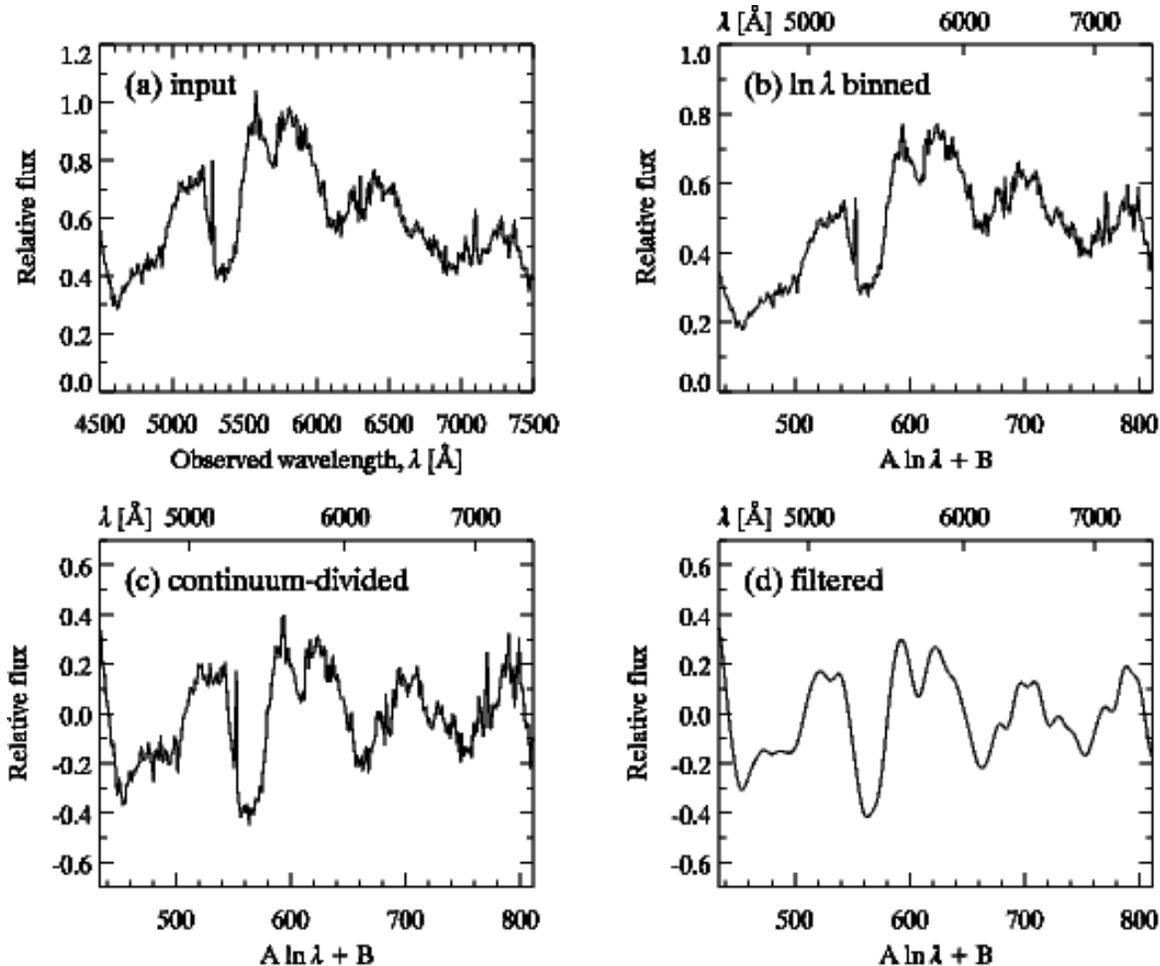}
\caption{Pre-processing the spectrum for \snid. ({\it a}) Spectrum of the
  SN~Ia SN~2003lj at $z=0.417$
  \citep{Matheson/etal:2005}. ({\it b}) The result of mapping the 
  spectrum to $\ln \l$ coordinates, with $(\l_0, \l_1, N) =
  (2500$\,\AA$, 10000$\,\AA$, 1024)$. ({\it c})
  A 13-point spline has been divided out and the result normalized to
  zero mean flux. ({\it d}) A bandpass filter with
  $(k_1,k_2,k_3,k_4)=(1,4,25,102)$ has been applied to the
  spectrum. 
\label{Fig:lnlambda}}
\end{figure*}

As already mentioned, the input and template spectra are binned on a
common logarithmic wavelength scale, characterized by $(\l_0,\l_1,N)$
(eq.~\ref{Eqn:dllog}). We show the result of mapping an input
supernova spectrum onto a logarithmic wavelength axis with $(\l_0,
\l_1, N) = (2500$\,\AA$, 10000$\,\AA$, 1024)$ in
Fig.~\ref{Fig:lnlambda}{\it a,b}. The size of a logarithmic
wavelength bin in this case is $d\l_{\rm ln} = \ln{(10000/2500)} /
1024 \approx 0.0014$, from eq.~\ref{Eqn:dllog}. So a shift by one bin
in $\ln \l$ space corresponds to a velocity shift of $d\l_{\rm ln}c
\approx 400$\kms. This is 1 order of magnitude less than the typical
width of a supernova spectral feature, Doppler-broadened by the $\sim
10000$\,\kms expansion velocity of the SN ejecta.

The next step in preparing the spectra for correlation analysis 
is continuum removal \citep{Tonry/Davis:1979}. 
For galaxy spectra, the continuum is well defined and is easily
removed using a least-squares polynomial fit. In
supernova spectra, however, the apparent continuum is ill-defined
due to the domination of bound-bound transitions in the total opacity
\citep{Pinto/Eastman:2000}. Dividing out a 13-point cubic 
spline fit to the spectrum (over the entire 2500--10000\,\AA\
wavelength interval) is akin to removing a {\it pseudo} continuum
from the supernova spectrum. We then subtract 1 from the resulting
spectrum and apply a normalization constant for the mean flux to equal
zero (Fig.~\ref{Fig:lnlambda}{\it c}). This effectively discards any
spectral color information (including reddening uncertainties and flux
mis-calibrations), and the correlation only relies on the {\it
  relative} shape and strength of spectral features in the input and
template spectra. We note that continuum division is also used by
  \citet{Jeffery/etal:2006} for measuring the goodness-of-fit between
  supernova spectra. We see below (\S~\ref{Sect:zt})
that the loss of color information has surprisingly little impact on
the redshift and age determination. Continuum removal also minimizes
discontinuities at each end of the spectrum, which would cause
artificial peaks in the correlation function. Further discontinuities
are removed by apodizing the spectra with a cosine bell ($\sim 5\%$ at
either end).

The final step is the application of a bandpass filter. While it is
actually applied at a later stage, directly to the correlation
function, we show its effect on the input spectrum in
Fig.~\ref{Fig:lnlambda}{\it d}. The goal is to remove low-frequency
residuals left over from the pseudo-continuum removal and
high-frequency noise components. Formally, the Fourier
transform of the normalized correlation function, $C(k)$
(eq.~\ref{Eqn:C(k)}), is multiplied by a real bandpass function
(so that no phase shifts are introduced) $B(k)$, such that

\begin{equation}
B(k) = \left\{ \begin{array}{ll}
0 & \textrm{for $k   \le k_1$ or $k \ge k_4$} \\
\onehalf \left[ 1 - \cos \left( \frac{k-k_1}{k_2-k_1}\right) \right] & \textrm{for $k_1 < k < k_2$} \\
1 & \textrm{for $k_2 \le k \le k_3$} \\
\onehalf \left[ 1 - \cos \left( \frac{k-k_3}{k_4-k_3}\right) \right]   & \textrm{for $k_3 < k < k_4$} \\
\end{array} \right.
\end{equation}

\begin{figure}	
\plotone{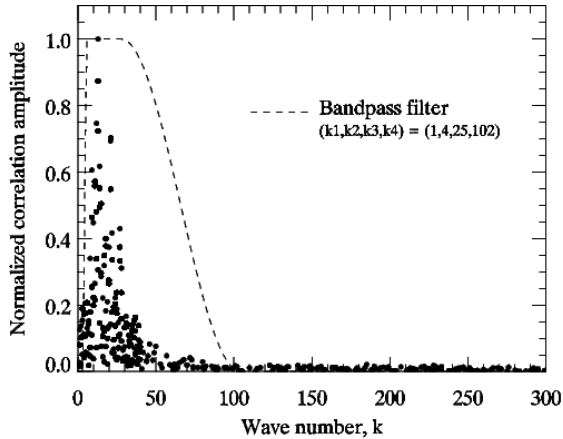}
\caption{Normalized amplitude of the Fourier transform of typical 
  unfiltered cross-correlation functions vs
  wavenumber, $k$. Note how most of the power is
  concentrated at low wavenumbers ($k \lesssim 50$), justifying our
  choice for the bandpass filter ({\it dashed line}). In this example
  the maximum correlation amplitude is at $k=13$, corresponding to a
  wavelength scale of $\sim 90$\,\AA.
\label{Fig:cfnamp}}
\end{figure}

The exact choices for the wavenumbers $(k_1,k_2,k_3,k_4)$ depend on
the size of each 
$k$ bin and on the spectral energy distribution of a supernova
spectrum. Supernova spectral lines have typical widths of
$\sim$100--150\,\AA, due to the large expansion velocities of the
ejecta ($\sim 10000$\,\kms). The mean 
size of a logarithmic wavelength bin with $(\l_0, \l_1, N) =
(2500$\,\AA$, 10000$\,\AA$, 1024)$ is $\sim 7.2$\,\AA, so a typical SN
line will have a width $w_{\rm line} \sim$ 14--21 logarithmic wavelength
bins. In Fourier space, most of the information will be at wavenumbers
$k = N / (2\pi \times w_{\rm line}) \approx$ 8--12. Since SN 
spectra consist of overlapping spectral lines \citep{Baron/etal:1996},
a typical SN feature may have a lower width ($\la 50$\,\AA). This
translates to $k \sim 25$, so most information is at wavenumbers less
than 25 and almost everything above wavenumber $k \sim 50$ is
noise. Also, low wavenumbers ($k \la 5$) contain information about
the low-frequency residuals from continuum removal. In
Fig.~\ref{Fig:cfnamp} we show the amplitude of the Fourier transform
of typical unfiltered correlation functions as a function of
wavenumber. As expected, most of the correlation power is at wave
numbers $k \sim$ 10--20, and virtually no information is contained in
wavenumbers $k > 50$.

\section{Redshift Estimate}\label{Sect:z}

In this section we introduce the correlation height-noise ratio
$r$ (\S~\ref{Sect:r}) and the spectrum overlap (${\rm lap}$) parameter (\S~\ref{Sect:lap}),
the product of which (the $r{\rm lap}$ quality parameter) conveys quantitative information about
the reliability of a cross-correlation redshift output by \snid. We
then briefly describe the redshift estimation
(\S~\ref{Sect:initrevz}) and associated error (\S~\ref{Sect:zerr}).

\subsection{The $r$-value\label{Sect:r}}

\begin{figure}	
\plotone{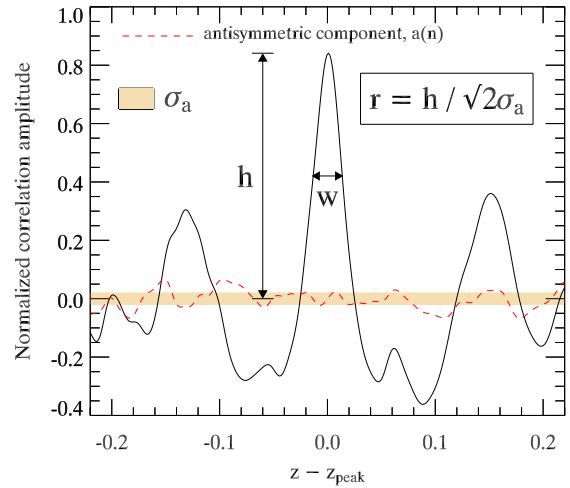}
\caption{Correlation height-noise ratio, $r$, is defined as the
  ratio of the height, $h$, of the highest peak in the normalized
  correlation function ({\it solid line}) to the rms of
  its antisymmetric component, $a(n)$ ({\it dashed line}), about the
  redshift corresponding to that peak ($z_{\rm peak}$). The width of
  the peak, $w$, is used to compute the redshift error ({\it
  see text for details}). {\it [See the electronic version of the
  Journal for a color version of this figure.]}
\label{Fig:rdef}}
\end{figure}

\citet{Tonry/Davis:1979} introduce the correlation height-noise ratio, $r$, to quantify the
significance of a peak in the normalized correlation function,
$c(n)$. It is defined as the ratio of the height, $h$, of the peak to
the rms of the antisymmetric component of $c(n)$,
$\sigma_a$, about the correlation redshift (Fig.~\ref{Fig:rdef}):

\begin{equation}
\label{Eqn:rdef}
r = \frac{h}{\sqrt{2} \sigma_a}.
\end{equation}

In order to compute $\sigma_a$, \citet{Tonry/Davis:1979} assume that
$c(n)$ is the sum of an auto-correlation of a template spectrum $t(n)$
with a shifted template spectrum $t(n-\delta)$ and of a random function
$a(n)$ that can distort the correlation peak:

\begin{equation}
\label{Eqn:tshift}
c(n) \equiv t(n) \star t(n-\delta) + a(n).
\end{equation}

\noindent
The first term on the right-hand side of eq.~\ref{Eqn:tshift} is
supposed to give a correlation peak of height $h=1$ at the exact
redshift (corresponding to a shift $\delta$ in logarithmic wavelength units),
while the second part can distort the peak. Since $t \star
t(n-\delta)$ is symmetric about $n=\delta$, the antisymmetric part of
$c(n)$ about $n=\delta$ equals the antisymmetric part of $a(n)$ about
$n=\delta$. We further assume that the symmetric part of $a(n)$ has
roughly the same amplitude as its antisymmetric part and that the
symmetric and antisymmetric parts of $a(n)$ are uncorrelated. In that
case, the rms of $a(n)$ is $\sqrt{2}$ times the rms of its
antisymmetric component.

\begin{figure}	
\epsscale{1.2}
\plotone{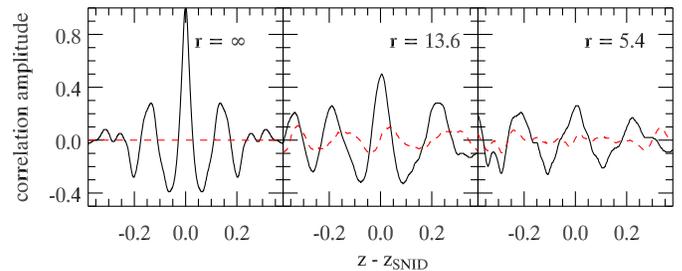}
\caption{Examples of perfect ({\it left}), good ({\it middle}), and
  poor ({\it right}) normalized correlation functions ({\it solid
  line}). The antisymmetric component of the correlation
  function about the \snid\ redshift ($z_{\rm \snid}$) is also shown
  ({\it dashed line}). {\it [See the electronic version of the
  Journal for a color version of this figure.]}
\label{Fig:cfnegs}}
\end{figure}

A perfect correlation will have a peak with $h=1$ at the exact
redshift, and $c(n)$ will be symmetric about $n=\delta$,
thus  $\sigma_a=0$ and so $r \longrightarrow \infty$
(Fig.~\ref{Fig:cfnegs}, {\it left panel}). Conversely, $r$
will be small ($r \lesssim 5$) for a spurious correlation peak
(Fig.~\ref{Fig:cfnegs}, {\it right panel}), and large ($r \gtrsim 10$) for a
significant peak, since $h$ will be close to 1 and $\sigma_a$ will be
small (Fig.~\ref{Fig:cfnegs}, {\it middle panel}).

\subsection{Spectrum Overlap\label{Sect:lap}}

In \snid, the correlation height-noise ratio $r$ alone does not
provide the estimator by which 
a correlation peak is deemed reliable. It is further weighted by the
overlap in $\ln \l$ space between the input spectrum and each of
the template spectra used in the correlation. In practice, the
template spectra are trimmed to match the wavelength range of the
input spectrum at the redshift corresponding to the correlation
peak. For an input spectrum with rest-frame wavelength range
$[\l_0,\l_1]$, the overlap in $\ln \l$ space, ${\rm lap}$, with each
template spectrum is in the range

\begin{equation}
\label{Eqn:laprange}
0 \le {\rm lap} \le \ln{\left( \frac{\l_1}{\l_0} \right)}.
\end{equation}

\noindent
Thus for an input and template spectra both overlapping the
rest-frame wavelength interval 3500--6000\,\AA,
${\rm lap}=\ln(6000/3500)\approx 0.54$.

The spectrum overlap parameter conveys important absolute information
about the quality of the correlation, complementary to the correlation
height-noise ratio $r$. Supposing a typical SN~Ia spectral feature has an FWHM of
$\Delta \l \approx 200$\,\AA\ at $\l \approx 5000$\,\AA, any correlation
with ${\rm lap}\lesssim \ln{(5400/5000)} \approx 0.08$ will be meaningless: {\it
any} feature will match any other at practically any redshift. Only
when a correlation has an associated ${\rm lap}$ that is several times
$\ln{(\Delta\l/\l)}$ can one rely on the redshift output by \snid. 

\begin{figure}	
\centering
\includegraphics[width=3.25in]{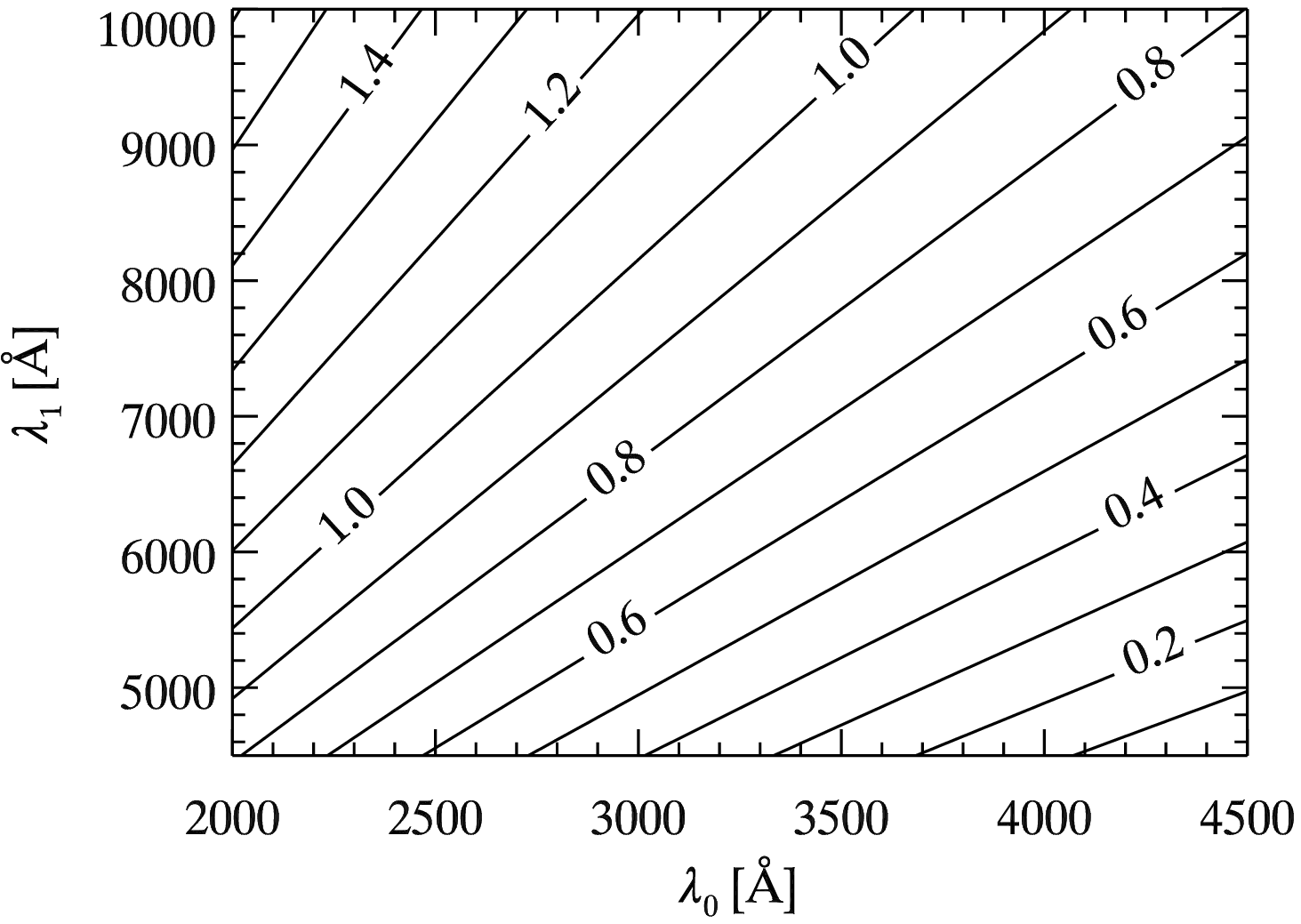} \\
\vspace{.35in}
\includegraphics[width=3.25in]{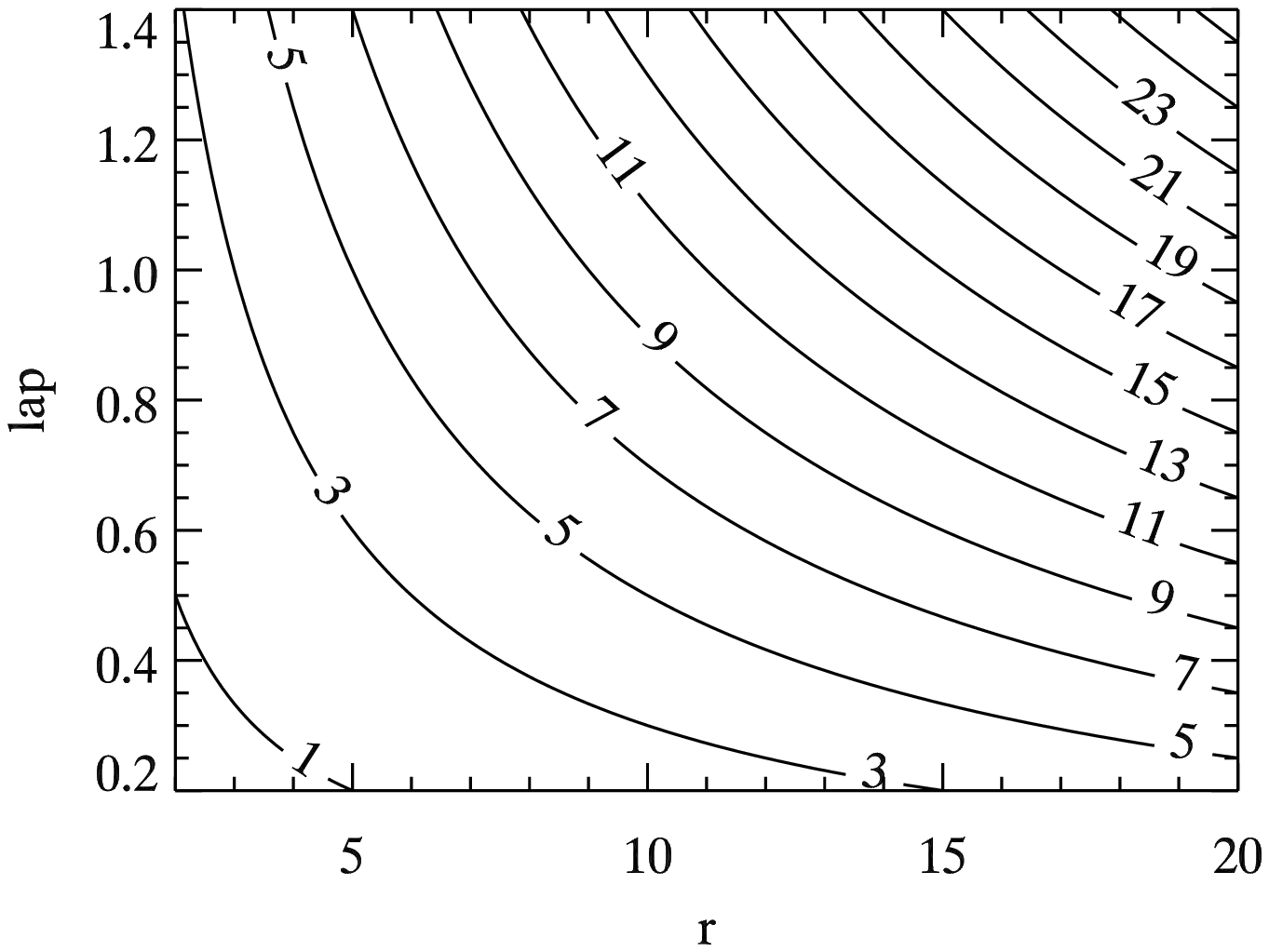}
\caption{{\it Top:} Contours of equal
  ${\rm lap}=\ln(\l_1/\l_0)$ for different rest-frame wavelength
  ranges $[\l_0,\l_1]$ of overlap between input and template
  spectra. We usually discard correlations with ${\rm lap} < 0.4$.
  {\it Bottom:} Contours of equal $r{\rm lap} = r \times {\rm lap}$ for a
  broad range of values for the correlation height-noise ratio
  ($r$) and spectrum overlap parameter (${\rm lap}$). We usually
  discard correlations with $r{\rm lap} < 5$ (and ${\rm lap} < 0.4$).
\label{Fig:rlapfigs}}
\end{figure}

In what follows, we usually discard correlation redshifts that have an 
associated ${\rm lap} < {\rm lap}_{\rm min} = 0.4$ and a quality parameter $r{\rm lap} = r
\times {\rm lap} < r{\rm lap}_{\rm min} = 5$. In Fig.~\ref{Fig:rlapfigs} we
  show contour plots for both the ${\rm lap}$ and $r{\rm lap}$ parameters, central
to the use of \snid.

\subsection{Initial and Revised Redshift Estimates\label{Sect:initrevz}}

For each template spectrum $t_i(n)$, we compute the correlation
function $c_i(n)=s(n)\star t_i(n)$. In general, $c_i(n)$ has many
peaks in redshift space (Fig.~\ref{Fig:rdef},\ref{Fig:cfnegs}). The
true redshift is most likely the one corresponding to the highest peak
in $c_i(n)$, although in poor signal-to-noise ratio cases some peaks
can distort or surpass the true redshift peak (Fig.~\ref{Fig:cfnegs},
{\it right panel}). In practice, \snid\ selects the 10 highest peaks
(labeled with index $j$) in $c_i(n)$ one by one and performs a fit
with a smooth function to determine the peak height and position,
$h_{ij}$ and $\delta_{ij}$, respectively. The corresponding redshift
is $z_{ij}=\exp(\delta_{ij}d\l_{\rm ln})-1$. The wavelength
regions of $s(n)$ and $t_i(n)$ that do not overlap at $z_{ij}$ are
trimmed, and, if the resulting spectrum overlap ${\rm lap} \ge {\rm lap}_{\rm min}$, a new
``trimmed'' correlation function, $c_{ij}(n)$, is computed, and the
corresponding correlation height-noise ratio ($r_{ij}$), spectrum
  overlap (${\rm lap}_{ij}$), and redshift ($z_{ij}$) are stored.

Once all templates have been cross-correlated with the
input spectrum, \snid\ computes an initial redshift, $z_{\rm
  init}$, based on an $r{\rm lap}$-weighted median of all $z_{ij}$. Each
redshift $z_{ij}$ is replicated $W_{ij}$ times according to the following
weighting scheme: 

\begin{equation}
W_{ij} = \left\{ \begin{array}{ll}
5 & \quad \textrm{for $\quad r{\rm lap}_{ij} >  6$} \\
3 & \quad \textrm{for $\quad r{\rm lap}_{ij} >  5$} \\
1 & \quad \textrm{for $\quad r{\rm lap}_{ij} >  4$} \\
0 & \quad \textrm{for $\quad r{\rm lap}_{ij}\le 4$,} \\
\end{array} \right.
\end{equation}

\noindent
If all $r{\rm lap}_{ij} \le 4$, $z_{\rm init}$ is set to 0.

\snid\ then computes a revised redshift based on the initial estimate, $z_{\rm init}$. The
input and template spectra (again labeled $i$) are  
trimmed such that their wavelength coverage coincides at $z_{\rm
  init}$. If the resulting spectrum overlap ${\rm lap}_{i} \ge {\rm lap}_{\rm min}$, a
second trimmed correlation function is computed and the
correlation height-noise ratio ($r_{i}$), spectrum overlap
  (${\rm lap}_{i}$), and redshift ($z_{i}$) corresponding
to the {\em highest} correlation peak are stored. The width
$w_{i}$ of the correlation peak is also saved and is used to
compute the redshift error (see next section). The revised redshift,
$z_{\snid}$, is computed as the non-$r{\rm lap}$-weighted median of all
redshifts $z_{i}$ that satisfy $r{\rm lap}_{i} \ge r{\rm lap}_{\rm min}$ with
${\rm lap}_{i} \ge {\rm lap}_{\rm min}$, with the additional requirement that
the individual redshifts $z_i$ do not differ significantly from the
initial redshift estimate: $|z_{i} - z_{\rm init}|
  < z_{\rm filt}$, where $z_{\rm filt} = 0.02$, typically.

\subsection{Redshift Error\label{Sect:zerr}}

One of the advantages of using the cross-correlation technique
for redshift determination is the ability to estimate the redshift
error, $\epsilon_z$. \citet{Tonry/Davis:1979} derive a formal
expression for $\epsilon_z$ based on the idea that spurious peaks
(positive and negative) in the antisymmetric component, $a(n)$,  of
the correlation function, $c(n)$, can distort the true correlation
peak. Obviously, $\epsilon_z$ is proportional to the number of peaks
in $a(n)$, and hence to the mean distance between peaks. Assuming
$c(n)$ and $a(n)$ have similar power spectra, the mean distance
between a peak in $c(n)$ and the nearest peak in $a(n)$ can be
estimated as $N/8B$ (\citealt{Tonry/Davis:1979}, their eq.~22), where
$N$ is the total number of bins and $B$ is the highest wavenumber at
the half-maximum point of the Fourier transform of $c(n)$ ($B \approx
25$ here; see Fig.~\ref{Fig:cfnamp}). One can then show that
(\citealt{Tonry/Davis:1979}, their eq.~24)

\begin{equation}
\epsilon_z = k_z \times \frac{1}{1+r},
\end{equation}

\noindent
where $k_z=N/8B$ ($\approx 5$ here) and $r$ is the correlation
height-noise ratio defined in
eq.~\ref{Eqn:rdef}. With the additional assumption of sinusoidal
noise in $c(n)$, \citet{Kurtz/Mink:1998} find $k_z=3w/8$, where $w$ is
the width of the correlation peak.

In practice, $k_z$ is calibrated using additional redshift measurements,
either through a different technique (e.g., 21\,cm measurements for
galaxy redshifts in \citealt{Tonry/Davis:1979}) or using the same
cross-correlation technique on two spectra of the same object
\citep{Kurtz/Mink:1998}. For supernova spectra, additional redshift
information potentially comes from narrow emission and absorption
lines in the host galaxy, while duplicate spectra of the same
supernova (at the same age) are not common
(Table~\ref{Table:sndb}). We find that including the spectrum overlap 
parameter (${\rm lap}$) yields a more robust estimator of the redshift error
(see also \S~\ref{Sect:zreserr}):

\begin{equation}
\label{Eqn:epsilonz}
\epsilon_z = k_z \times \frac{1}{1+r{\rm lap}},
\end{equation}

\noindent
with $k_z \approx$ 2--4$w$.

%
%
%


\section{The \snid\ Database}\label{Sect:sniddb}

\subsection{Nomenclature and Age Distribution}

\begin{figure}	
\plotone{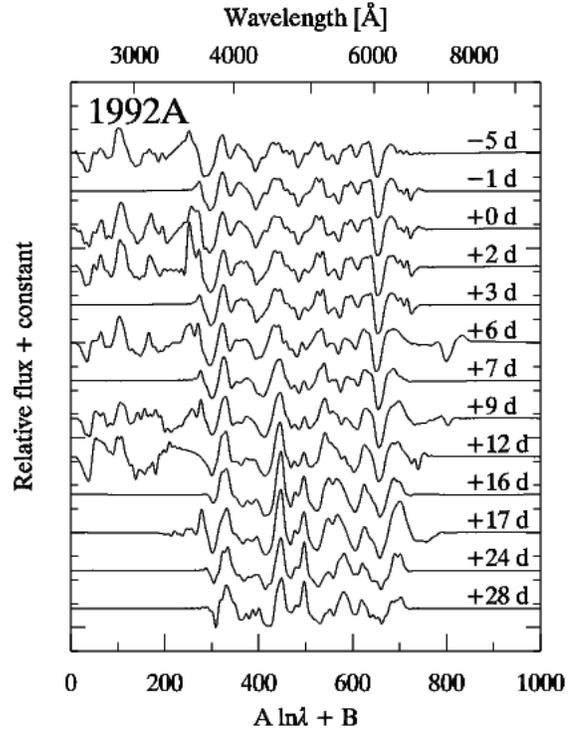}
\caption{Example of an \snid\ template, here the SN~Ia SN~1992A
  \citep{Kirshner/etal:1993}. We also indicate the age of each
  spectrum (in days from $B$-band maximum). Discontinuities in
  flux around $\sim3500$\,\AA\ are due to a calibration mismatch
  between the UV and optical components of the spectrum.
\label{Fig:doit}}
\end{figure}

The current \snid\ spectral database comprises 
879 spectra of 65 SNe~Ia, 322 spectra of 19 SNe~Ib/c, and 353 spectra of
10 SNe~II (Table~\ref{Table:sndb}). The spectra are drawn from public archives 
(SUSPECT\footnote{{\tt
    http://bruford.nhn.ou.edu/$\sim$suspect/index1.html}} and the CfA
Supernova Archive\footnote{{\tt
http://www.cfa.harvard.edu/supernova/SNarchive.html}}) 
and from the CfA Supernova Program \citep{Matheson/etal:2007}. The spectra are
chosen to have high signal-to-noise ratio (typically $\gtrsim 10$ per
\AA) and to span a sufficiently large optical rest frame wavelength
range ($\l_{\rm min} \le 4000$\,\AA; $\l_{\rm max} \ge 6500$\,\AA) to
include all the identifying features of SN spectra. We
remove telluric features in all the spectra, either using the
well-exposed continua of spectrophotometric standard stars for the
CfA data \citep{Wade/Horne:1988,Matheson/etal:2000}, or using a
simple linear interpolation over the strong A- and B-bands. We show
the full suite of spectra for the local SN~Ia spectral template
SN~1992A \citep{Kirshner/etal:1993}, which also includes some UV data
from the {\it Hubble Space Telescope} at some epochs, shifted to zero
redshift in Fig.~\ref{Fig:doit}.

While we have included all the supernovae available to us for which
there are a large number of epochs of spectroscopy, there are still
many more ($> 1000$) supernovae for which there are only one to two
epochs of spectroscopy that we have yet to include in the
database. We also include spectra of galaxies, active galactic nuclei, stars (including
variable stars, such as luminous blue variables), and novae. This can
be particularly useful when trying to weed out contaminants
from large surveys of high-redshift supernovae
(cf. \citealt{Matheson/etal:2005}).

\begin{figure}
\epsscale{1.1}
\plotone{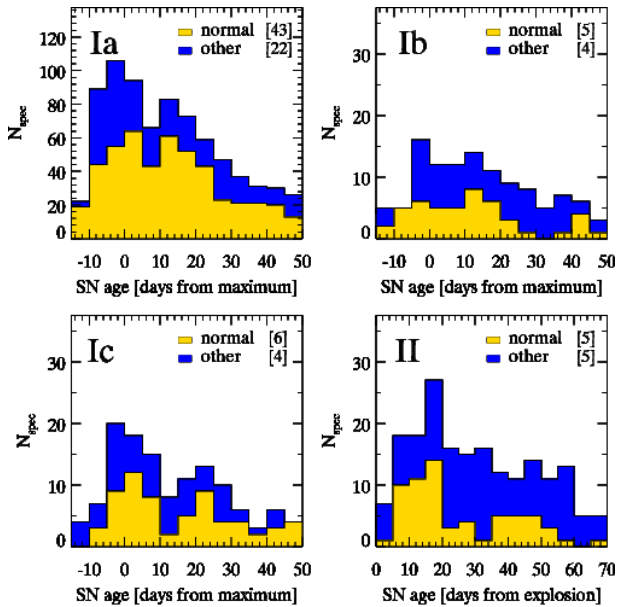}
\caption{Age distributions of \snid\ templates for supernovae of
  different types. The number of supernovae corresponding to a given
  type is indicated in square brackets. Note the larger
  ordinate range for SNe~Ia. For SNe~II, the
  age is given in days from the estimated date of explosion as
  opposed to days from maximum in a specific band. The transitional
  SNe~IIb are included in both Type Ib and Type II
  histograms. {\it [See the electronic version of the Journal for a
  color version of this figure.]}
\label{Fig:thist}}
\end{figure}

We show the age distribution of \snid\ supernova templates for the
main supernova types (Ia, Ib, Ic, II) in Fig.~\ref{Fig:thist}. 
For each type, we show the age distribution of ``normal''
representatives of that type, as well as spectra that show deviations
from the latter (in the ``other'' category). We note that this
division is somewhat qualitative and relies on the identification by
eye of certain characteristic spectroscopic features in the spectra
\citep{Filippenko:1997}. We are currently working on a statistical
scheme to separate our template spectra in these various categories
(see also Fig.~\ref{Fig:meanspec}). The nomenclature for the
different supernova types and their associated subtypes is given in
Table~\ref{Table:stype}. From Fig.~\ref{Fig:thist}, it is clear that
the age distribution of the \snid\ templates is not uniform, and
even bi-modal for SNe~Ia. This potentially introduces age
``attractors'' that could in principle bias the age and
redshift determination (although see \S~\ref{Sect:zt}). The 
fact that there are more SN~Ia templates than SNe~Ib, Ic, and II {\it
  combined} also leads to a type ``attractor,'' with the risk for
low-S/N spectra to be preferentially classified as SN~Ia, regardless
of their type (see \S~\ref{Sect:type}). 

\begin{deluxetable}{ l | l l l l }
\tabletypesize{\scriptsize}
\tablenum{2}
\tablewidth{0pt}
\tablecaption{Supernova types and subtypes\label{Table:stype}}
\tablehead{
\colhead{Type} &
\colhead{Ia} & 
\colhead{Ib} & 
\colhead{Ic} &
\colhead{II}
}
\startdata
``normal'' & Ia-norm & Ib-norm & Ic-norm  & II-norm (IIP) \\ \hline
           & Ia-pec  & Ib-pec  & Ic-pec   & II-pec        \\
``other''  & Ia-91T  & IIb     & Ic-broad & IIL           \\
           & Ia-91bg &         &          & IIn           \\
           &         &         &          & IIb           
\enddata
\tablecomments{
``norm'' and ``pec'' refer to ``normal'' and ``peculiar'' subtypes
  of the corresponding type; see Table~\ref{Table:sndb} for specific
  examples. ``Ic-broad'' is used to identify broad-lined SNe~Ic
(``hypernovae''), some of which are associated with Gamma-Ray
Bursts. The transitional Type IIb supernovae are included in both
Type Ib and Type II categories.
}
\end{deluxetable}

The execution time of \snid\ scales linearly with the number of
templates\footnote{Execution time $t_{\rm exec} \approx 6s ({\rm cpu}/{\rm 2.86
GHz})(N_{\rm temp}/1000)$, where $N_{\rm temp}$ is the number of 
templates.} and is remarkably low compared with $\chi^2$-based
methods (see \S~\ref{Sect:comparison})--- although see
  \citet{Rybicki/Press:1995} for fast statistical methods that can
  compete with the cross-correlation technique. It is trivial to
include large spectroscopic data sets--- such as those from
the CfA Supernova Group (for example, 431 spectra of 32 SNe~Ia,
included in the present \snid\ database, will soon be published by
\citealt{Matheson/etal:2007}).

\subsection{Intrinsic Spectral Variance}

\begin{figure*}
\plottwo{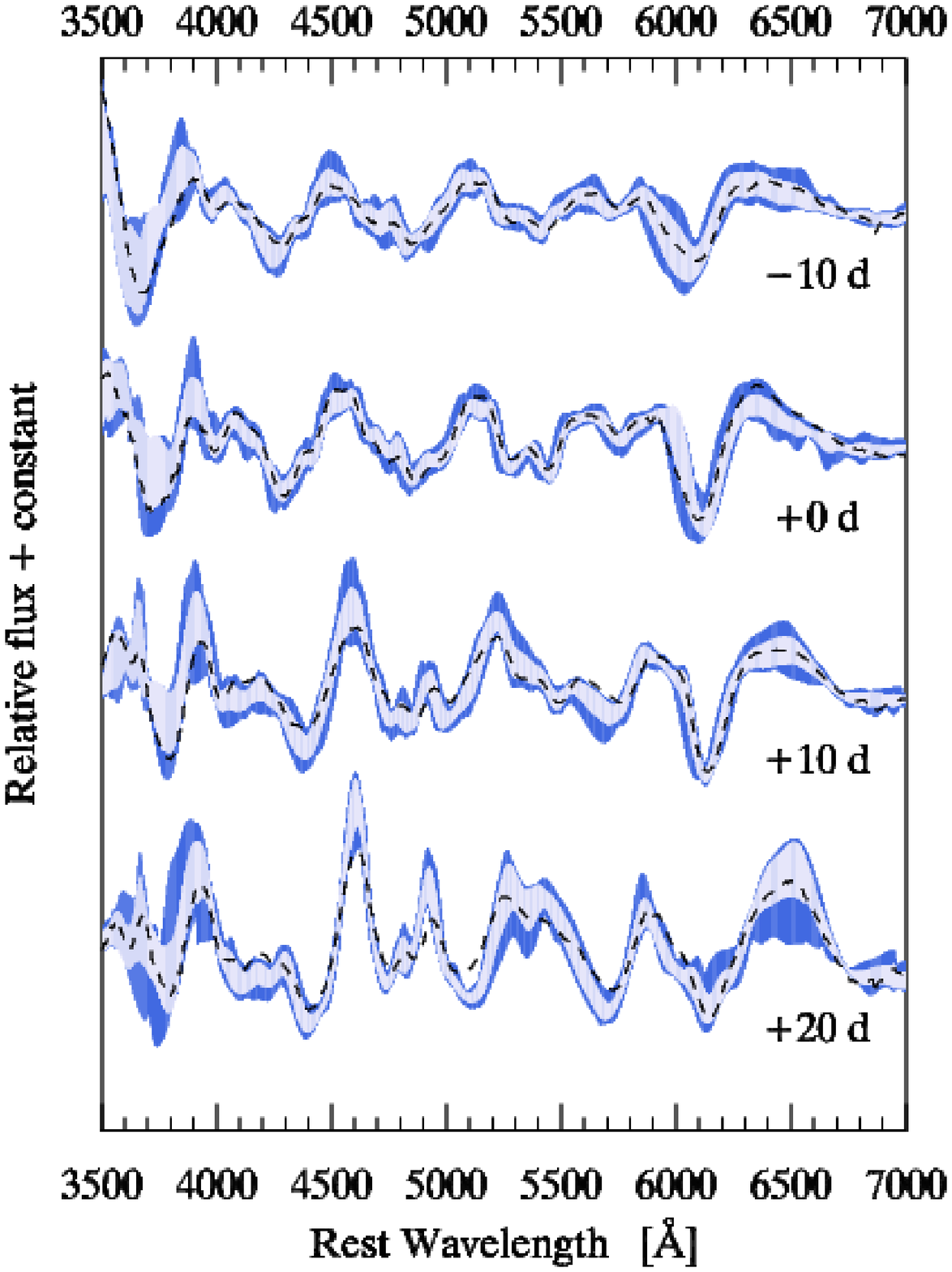}{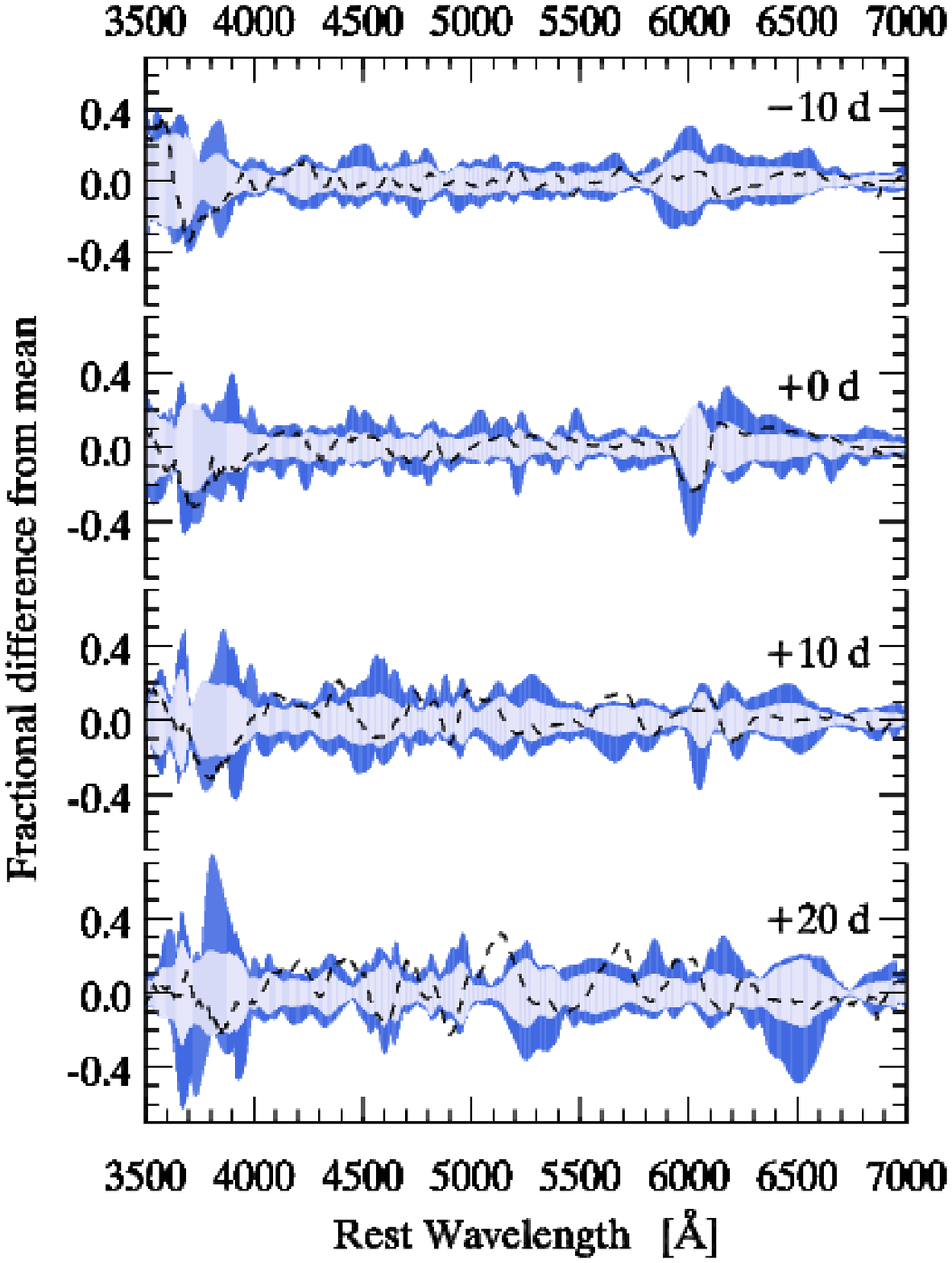}
\caption{
{\it Left:} Standard ({\it light gray}) and maximum ({\it dark gray})
deviation from the mean spectrum for all Ia-norm templates, at
four different ages. We overplot the corresponding Nugent template
at each age ({\it dashed line}). All spectra have been pre-processed
in the same way as any \snid\ template (\S~\ref{Sect:preproc}).
{\it Right:} Fractional difference from the mean Ia-norm
spectrum. We also show the ratio of the mean spectrum to the
corresponding Nugent template ({\it dashed line}). {\it [See the
    electronic version of the Journal for a color version of this
    figure.]}
\label{Fig:meanspec}}
\end{figure*}

In Fig.~\ref{Fig:meanspec} we show the standard and maximum deviation
from the mean spectrum of all ``Ia-norm'' templates at $-10$, +0, +10,
and +20 days from maximum light. One clearly sees the rapid variation
of SN spectra around maximum light, but also the change in intrinsic
scatter with age. For instance, the intrinsic spread in the strength
and position of the defining \sitwo\ \l6355 feature (which causes the
deep blueshifted absorption around $\sim 6100$\,\AA) decreases between
$-10$ to $+10$ days from maximum light. At +20 days, the scatter is
large in that wavelength region due to increasing contribution from
other ions (mainly \fetwo).

The residual variation about the mean spectrum
(Fig.~\ref{Fig:meanspec}, {\it right panel}) shows that normal SNe~Ia
are typically within 10\%--20\%\ from the mean spectrum, although
deviations greater than 40\%\ are seen at certain wavelength intervals
(again depending on the age). The fact that all Ia-norm
template spectra are within two standard deviations from a mean
spectrum suggests a possible statistical classification scheme to
differentiate normal SNe~Ia from the other Ia subtypes. With more
data, it is in principle possible to do this more reliably for SNe~Ia,
as well as other supernova types.

The intrinsic variation of the Ia-norm templates points to the
inadequacy of describing a given SN subtype with a single
representative template, unless the latter includes this
variance explicitly. Past attempts to create grids of such template
spectra, such as those presented by
\citet{Nugent/Kim/Perlmutter:2002}, do not account for the variability
within a given SN type at a given age. We show the corresponding
Nugent template (ver. 1.2) at each age in Fig.~\ref{Fig:meanspec} ({\it
  dashed line}). While most of the Nugent template is included within
the standard deviation from the mean spectrum in our database, there
are also significant deviations. We do note, however, that the
comparison is somewhat misleading since
\citet{Nugent/Kim/Perlmutter:2002} had less data available to them for
the elaboration of these templates. Nevertheless, we have tested their
use in the \snid\ spectral database, but have found them to lead to
systematic errors in both the redshift and age determination.


\section{Accuracy of Redshift and Age Determination}\label{Sect:zt}

We use a simple simulation to test the accuracy of \snid\ in
determining the redshift and age of a supernova spectrum. Here we
focus on normal SNe~Ia since they are the most
represented in our spectral database, although the conclusions of
this section are qualitatively valid for all other supernova
types. Even though normal SNe~Ia form a homogeneous class, the
spectra reveal intrinsic variations at any given age that affect
directly the redshift and age determination. The redshift precision
depends primarily on the typical width of a spectral feature
(decreasing from broad-lined SNe~Ic to SNe~IIn), which
affects the width of the correlation peak (see
Fig.~\ref{Fig:rdef}). The redshift accuracy depends primarily on the
intrinsic variation of line positions at a given age. The age
determination strongly correlates with the redshift determination
(\S~\ref{Sect:ztcovar}), and depends on how quickly the SN spectra
evolve at a given age.

\subsection{Presentation of the Simulation\label{Sect:simulation}}

In this simulation, each Ia-norm spectrum in the \snid\ database
(cf. Table~\ref{Table:sndb}) is correlated with all other Ia-norm
spectra, except for those corresponding to the input supernova (to
ensure unbiased results). We require all spectra used in the
simulation to include the rest-frame wavelength interval
3700--6500\,\AA and to have an age (in days from $B$-band
maximum, hereafter $t_B$) $-10 \le t_B \le +20$.

We show the simulation parameters in Table~\ref{Table:simparams}. The
input spectrum is first redshifted to $z$ by simply multiplying
the wavelength axis by $(1+z)$. We then ``contaminate'' the input
supernova spectrum with galaxy light (up to 50\% of the total flux),
using the elliptical and Sc galaxy templates of
\citet{Kinney/etal:1996}, and add noise (both random Poisson noise and
sky background) to reproduce the range of typical signal-to-noise
ratio of SN spectra at the simulation redshifts, when observed
with 6.5--10\,m-class telescopes (e.g., VLT, Keck, Gemini, Magellan)
used in cosmological SN~Ia surveys. Note that we do not scale
the input spectral flux to match a given simulation redshift, as
\snid\ normalizes the input and template spectra in a
similar fashion (Fig.~\ref{Fig:lnlambda}).

\begin{deluxetable*}{ll}
\tabletypesize{\scriptsize}
\tablenum{3}
\tablewidth{0pt}
\tablecaption{Simulation parameters\label{Table:simparams}}
\tablehead{
\colhead{Parameter} & 
\colhead{Range}
}
\startdata
Redshift, $z$ & $0.1 \le z \le 0.7$ \\
Galaxy contamination fraction, $f_{\rm gal}$ & $0 \le f_{\rm gal} \le 0.50$ \\
Signal-to-noise ratio, S/N (per 2\,\AA) & $1 \le {\rm S/N} \le 15$ \\
Age (days from $B$-band maximum), $t_B$ & $-10 \le t_B \le +20$ \\
Minimum rest frame wavelength coverage, $\l_{\rm rest}$ (\AA) & $3700 \le \l_{\rm test} \le 6500$ \\
Observed wavelength range, $\l_{\rm obs}$ (\AA) & $4000 \le \l_{\rm obs} \le 9000$ \\
\enddata
\end{deluxetable*}

We restrict the {\it observed} wavelength range over which \snid\
computes the correlation to $4000 \le \l_{\rm obs} [$\AA$]\le 9000$,
to mimic the coverage of the FORS1 optical spectrograph mounted on the
VLT. We have not studied the impact of a change in this wavelength
range on the redshift or age determination. Furthermore, we force
\snid\ to only consider correlation redshifts in the interval
$[0,1]$. For each correlation, we record the template name, type,
subtype, and age; the correlation redshift and its associated 
  correlation height-noise ratio ($r$) and spectrum overlap parameter
  (${\rm lap}$); and the width $w$ of the correlation peak (to estimate the 
redshift error).

To study the effects of constraints on redshift and age, we run
\snid\ three times on the input spectrum: once with no constraints; a
second time with a flat constraint on redshift ($\pm 0.01$), and a third
time with a flat constraint on age ($\pm 3$ days). We note that the
distribution of redshift residuals is remarkably Gaussian
(\S~\ref{Sect:zreserr}), and we are currently implementing Gaussian
priors in \snid. A total of 4 billion correlations were computed
with \snid\ for this simulation, in just under 70 CPU hr.

\subsection{Redshift Residuals and Redshift Error\label{Sect:zreserr}}

\begin{figure}
\epsscale{1.2}
\plotone{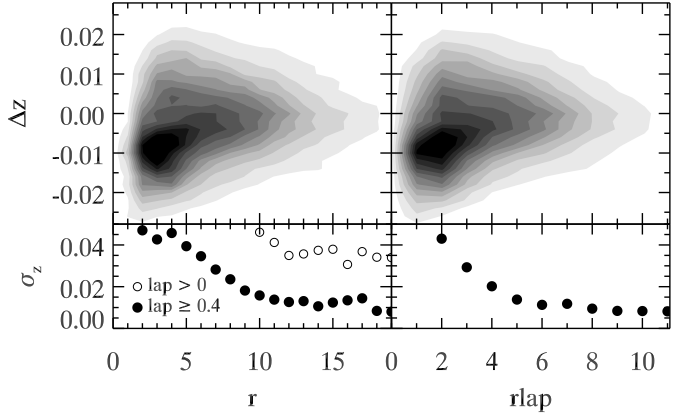}
\caption{{\it Top:} 2D histograms of redshift
  residuals vs. the correlation height-noise ratio $r$
  ({\it left panel}) and the $r{\rm lap}$ quality parameter (with ${\rm lap} \ge
  0.4$; {\it right panel}), with the following parameters: $0.3 \le z \le
  0.5$, $-5 \le t_B \le +15$, $2 \le$ S/N (per
  2\,\AA)$\le 10$. The linear gray scale reflects the number of points
  in a given 2D bin (the more points the darker). 
  {\it Bottom:} Standard deviation, $\sigma_z$, of
  redshift residuals in $r,r{\rm lap}$ bins of size unity. For the
  $\sigma_z(r)$ curve ({\it filled circles, bottom left}), we show the
  effect of additionally requiring that ${\rm lap} \ge 0.4$ ({\it open
  circles}).
\label{Fig:zsigrlap}}
\end{figure}

We show the distribution of redshift residuals, $\Delta z$, versus
the $r{\rm lap}$ quality parameter in the top-right panel of
Fig.~\ref{Fig:zsigrlap}, for input parameters  $0.3 \le z \le
0.5$, $-5 \le t_B \le +15$, and $2 \le$ S/N (per 2\,\AA)$\le
10$. The residuals are shown as a two-dimensional (2D) histogram, with 
a linear gray-scale scheme reflecting the number of points in a given
$(\Delta z,r{\rm lap})$ bin. We only show correlations for which the overlap
between input and template spectra ${\rm lap} \ge 0.4$. For good
correlations ($r{\rm lap} \gtrsim 5$), the distribution of redshift
residuals is a Gaussian centered at $\Delta z = 0$. In the bottom right
panel, we show the standard deviation of redshift residuals,
$\sigma_z$, in $r{\rm lap}$ bins of size unity. For $r{\rm lap} \gtrsim 5$, we
have a typical error in redshift of order $\sigma_z \lesssim 0.01$.

\begin{figure}
\epsscale{1.1}
\plotone{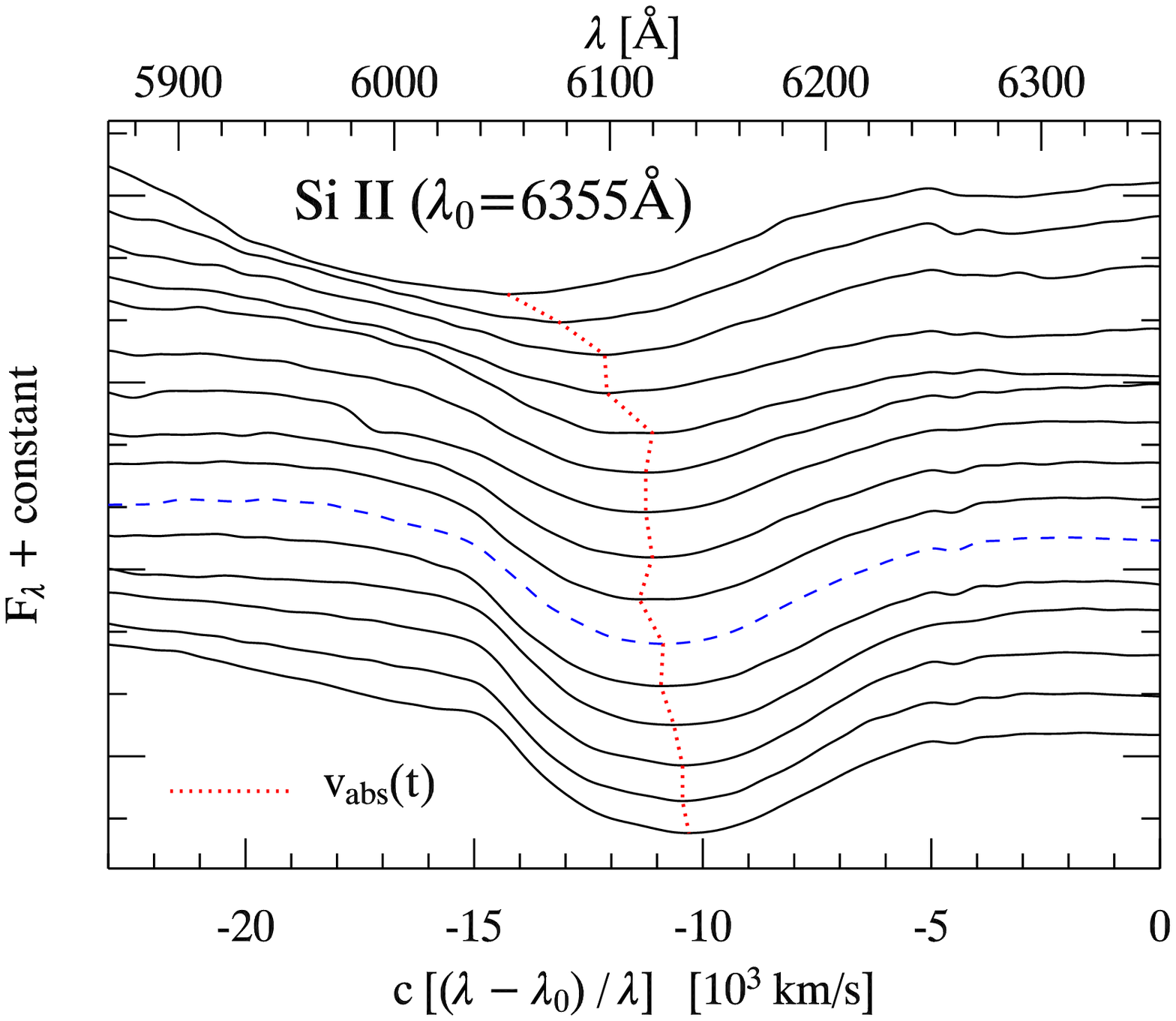}
\caption{Evolution of the blueshifted \sitwo\ \l6355 absorption
  profile in the SN~Ia SN~1994D
  \citep{Hoeflich:1995,Patat/etal:1996} between $-11$ and +7 days
  from $B$-band maximum. The dotted line shows the velocity location
  of the locus of maximum absorption, $v_{\rm abs}$. We highlight the
  \sitwo\ profile at maximum light ({\it dashed line}). Over the course
  of 18 days, the locus of maximum absorption shifts redward in
  wavelength by $\sim 80$\,\AA, corresponding to $\sim 4000$\,\kms\ in
  velocity. Note the more rapid evolution of $v_{\rm abs}$ before
  maximum light. {\it [See the electronic version of the Journal for a
  color version of this figure.]}
\label{Fig:vabst}}
\end{figure}

For poor correlations ($r{\rm lap} \lesssim 3$) there is a
concentration of points around $\Delta z \approx -0.01$. 
This is an artifact of the {\it pseudo}-continuum removal, which
enhances the contrast between emission peaks and
absorption troughs in the input and template spectra and biases
poor correlations to later ages. In this simulation, many input
spectra at maximum are attracted to $\sim +10$ days,
where the position of SN spectral
features has shifted redward in wavelength due to the expansion of the
supernova envelope (Fig.~\ref{Fig:vabst}). The template needs to be
shifted less in $\ln \l$ space to match the redshift of the input
spectrum, which leads to an under-estimation of the redshift by $\sim
0.01$. This corresponds to a combination of the typical velocity shift
in SN~Ia absorption features from maximum to $\sim 10$ days past
maximum, and the spread of these velocities at a given age ($\sim
3000$\kms; see \citealt{Benetti/etal:2005,Blondin/etal:2006a}).
We note that this artifact has no impact on correlations with $r{\rm lap}
> 3$.

In the left panels of Fig.~\ref{Fig:zsigrlap}, we show the same
distribution of $\Delta z$, this time only as a function of the
  correlation height-noise ratio, $r$ (note
the change in the abscissa range). To first order, the 
2D histogram of redshift residuals looks remarkably
similar to that as a function of $r{\rm lap}$ (Fig.~\ref{Fig:zsigrlap}, {\it
  right panels}), with again a concentration of
points around $\Delta z \approx -0.01$ for low values of $r$. However,
the variation of $\sigma_z$ with $r$ ({\it bottom left panel, filled
  circles}) gives a different picture: the lack of constraint on ${\rm lap}$
causes in some cases a mis-estimate of the redshift, at all $r$,
thereby greatly biasing $\sigma_z$ to higher values ($\sigma_z > 0.03$, for
all $r$). Requiring that ${\rm lap} \ge 0.4$ leads to a significant improvement
({\it open circles}), with $\sigma_z \lesssim 0.01$
for $r \gtrsim 10$. It is therefore imperative to consider the overlap
between the input and template spectra to yield accurate supernova
redshifts with the cross-correlation technique.

\begin{figure}
\epsscale{1.2}
\plotone{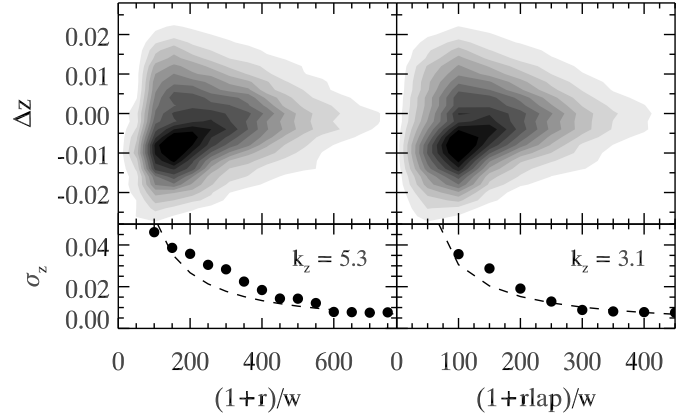}
\caption{Same as Fig.~\ref{Fig:zsigrlap}, except the abscissae now
  correspond to $(1+r,r{\rm lap})/w$, where $w$ is the width of the
  correlation peak. A fit to the binned $\sigma_z$ distributions ({\it
  bottom}) yields the value for $k$ used in estimating the error
  (eq.~\ref{Eqn:epsilonz}). In both panels, only
  correlations with ${\rm lap} \ge 0.4$ are shown.
\label{Fig:zsigrlapw}}
\end{figure}

The formal redshift error, $\epsilon_z$, is proportional to
$w/(1+r{\rm lap})$ (eq.~\ref{Eqn:epsilonz}), $w$ being the width of the
correlation peak (Fig.~\ref{Fig:rdef}). We illustrate the
determination of the constant of 
proportionality, $k_z$, in Fig.~\ref{Fig:zsigrlapw}, where we show the
same 2D histograms of redshift residuals $\Delta z$, this
time as a function of $(1+r)/w$ ({\it left panels}) and $(1+r{\rm lap})/w$ ({\it
  right panels}). A best fit to the $\sigma_z$ curves in the bottom panels
yields a value for $k_z$: 5.3 for $(1+r)/w$, and 3.1 for
$(1+r{\rm lap})/w$. Only correlations with ${\rm lap} \ge 0.4$ are shown. As in
Fig.~\ref{Fig:zsigrlap}, the product of the $r$-value and the
overlap yields a more robust error estimator than the
$r$-value alone. In what follows we study variations of redshift and
age determinations using \snid\ only as a function of the $r{\rm lap}$
quality parameter, with ${\rm lap} \ge 0.4$.

In principle, $k_z$ needs to be evaluated for every template spectrum in
the database, through either internal or external comparisons (as done
for galaxy spectral templates in
\citealt{Tonry/Davis:1979,Kurtz/Mink:1998}). While this is impractical
for supernova spectra--- there are few duplicate spectra of the same
supernova at a given age (Table~\ref{Table:sndb}), we have
computed $k_z$ using subsets of templates used in our simulation, as
well as for other supernova types, and have found that $k_z$ is
typically in the range $2 \lesssim k_z \lesssim 4$, with $k_z\approx 3$
being the median value.

\begin{figure}
\epsscale{1.2}
\plotone{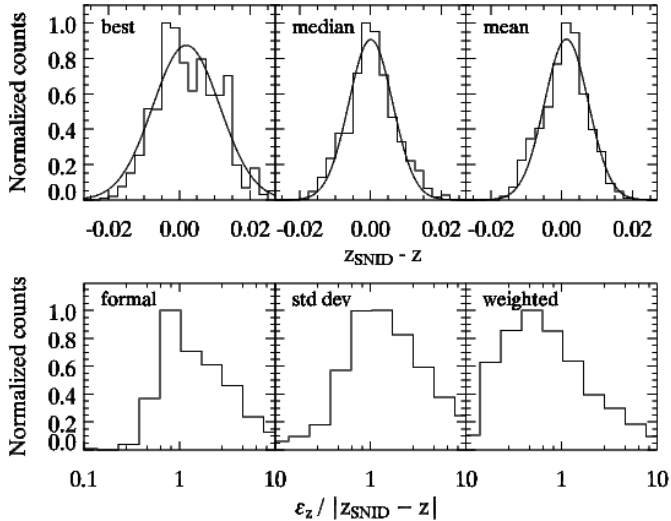}
\caption{{\it Top:} Normalized distributions of redshift
  residuals, when the \snid\ redshift is assumed to be the redshift
  of the best-match template ({\it left}), the median redshift ({\it
  middle}), and the $r{\rm lap}$-weighted mean redshift ({\it right}). The
  distribution of median redshifts is the most consistent with a
  normal distribution. {\it Bottom:} Normalized distributions of
  the ratio of the absolute redshift residual (corresponding to the
  different redshift estimators in the top panel) to the redshift
  error, $\epsilon_z$, estimated in different ways (see text for
  details). 
\label{Fig:deltazhist}}
\end{figure}

The above holds for a {\it single} spectral template; to use \snid\ to
its full capacity, we need to combine redshifts for {\it all} templates
for which the $r{\rm lap}$ quality parameter is greater than a certain cutoff (generally, $r{\rm lap}
\ge r{\rm lap}_{\rm min} = 5$). In \S~\ref{Sect:initrevz}, we favored the
non-$r{\rm lap}$-weighted median of all correlation redshifts with $r{\rm lap} \ge
r{\rm lap}_{\rm min}$ as being ``the'' \snid\ redshift, but did not justify
this. In Fig.~\ref{Fig:deltazhist} ({\it top panels}), we show
distributions of \snid\ redshift residuals, when the \snid\ redshift
is taken to be the redshift corresponding to the highest $r{\rm lap} \ge 5$
value (``best;'' {\it left panel}); the median ({\it middle panel}) or
$r{\rm lap}$-weighted mean ({\it right panel}) of all redshifts with $r{\rm lap} \ge
5$. Both the median and mean distributions are consistent with a
Gaussian distribution, with the median redshift providing a slightly
better match. This is expected since the use of the median redshift
guards us from systematic errors produced by spurious or ill-defined
correlation peaks for some templates. The distribution of ``best''
redshift residuals is broader and non-uniform. We therefore consider
the median redshift to provide the best estimate.

In the bottom panels of Fig.~\ref{Fig:deltazhist} we show the normalized
distributions of the ratio of the absolute redshift residual
(corresponding to the different redshift estimators in the top
panel) to the redshift error, $\epsilon_z$, estimated in different
ways. A ratio equal to or above unity indicates that the actual redshift is
consistent with the \snid\ redshift within the estimated error, while a ratio
below unity indicates that the error is under-estimated. For a good error
estimator, we expect those distributions to peak at a ratio near
unity, with a long tail to higher ratios and a sharp drop below
 unity. Such is the case for the formal redshift error
(eq.~\ref{Eqn:epsilonz}) associated with the ``best'' redshift ({\it
  left panel}). It is not obvious which error to associate with the median
redshift. We found that the standard deviation of all correlation
redshifts with $r{\rm lap} \ge 5$ provided a satisfactory estimate of the
error ({\it middle panel}). This same estimator was used by
\citet{Matheson/etal:2005} for high-$z$ SN~Ia spectra from the ESSENCE
survey. The error in the $r{\rm lap}$-weighted mean ({\it right panel}), on the
other hand, systematically underestimates the true redshift error by
a factor of $\sim 3$.

\subsection{Age Residuals}

Unlike redshift, the supernova age is not (and cannot be)
a free parameter in \snid, as it is a discrete variable tied in with a
specific spectral template. Nevertheless, since the 
cross-correlation technique relies solely on the relative strengths
and position of broad spectroscopic features, which themselves are a
strong function of the supernova age
(Figs.~\ref{Fig:doit},\ref{Fig:meanspec}, \&~\ref{Fig:vabst}), we
expect a strong correlation between the $r{\rm lap}$ quality parameter and the age
residual, $\Delta t$, between input and template spectra.

\begin{figure}
\epsscale{1.1}
\plotone{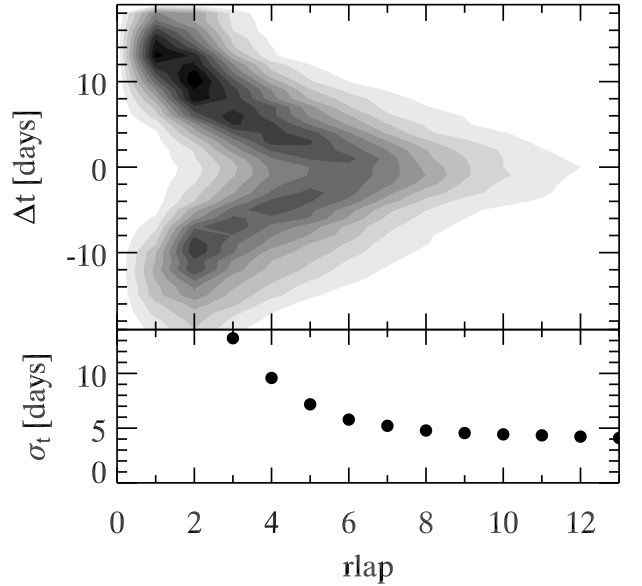}
\caption{{\it Top:} 2D histogram of age
  residuals vs. the $r{\rm lap}$ quality parameter (with ${\rm lap} \ge 0.4$), with the same
  parameters as in Fig.~\ref{Fig:zsigrlap}. The gray scale reflects the
  number of points in a given 2D bin (the more points the darker). 
  {\it Bottom:} Standard deviation, $\sigma_t$, of age
  residuals in $r{\rm lap}$ bins of size unity. For $r{\rm lap} \gtrsim 6$,
  $\sigma_t \lesssim 5$ days. 
\label{Fig:tsignf}}
\end{figure}

We show the distribution of age residuals versus $r{\rm lap}$ in
Fig.~\ref{Fig:tsignf} ({\it top panel}), where the gray scale has the
same meaning as in the previous 2D histograms. For $r{\rm lap}
\gtrsim 6$, the distribution of age residuals is a Gaussian centered
at $\Delta t = 0$. In the bottom panel, we show the standard deviation
of age residuals, $\sigma_t$, in $r{\rm lap}$ bins of size unity. For
$r{\rm lap} \gtrsim 6$, we have a typical error in age of order
$\sigma_t \lesssim 5$ days.

The most striking feature in the gray scale of Fig.~\ref{Fig:tsignf} is
the near absence of points around $\Delta t=0$ for low values of
$r{\rm lap}$. For poor correlations, the age is systematically
mis-estimated, with a tendency to overestimate the age by $\sim10$
days. Again, this is an artifact of the {\it pseudo-}continuum
removal, which causes many maximum-light spectra to correlate with
$\sim +10$-day templates (see \S~\ref{Sect:zreserr}).
It is also due to the
nature of the supernova evolution, as the spectra evolve more rapidly
around maximum light than they do around 10 days past maximum
(Fig.~\ref{Fig:meanspec}), increasing the likelihood of correlations
with templates at these ages.

There is no formal estimator for the age error. We have examined the
distribution of age residuals above a certain $r{\rm lap}$ cutoff 
(cf. Fig.~\ref{Fig:deltazhist} for redshift) and find the median age of
all templates with $r{\rm lap} \ge 5$ to be a good estimate of the spectral
age. However, the standard deviation of all template ages with $r{\rm lap}
\ge 5$ tends to systematically overestimate the age error by $\sim
20\%$.

\subsection{Covariance Between Redshift and Age\label{Sect:ztcovar}}

\begin{figure}
\epsscale{1.1}
\plotone{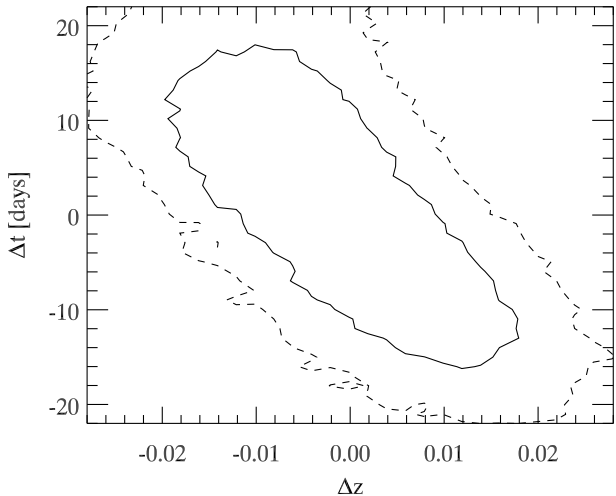}
\caption{Age residuals vs. redshift residuals, illustrating the
  covariance between the two quantities. We show the $1\sigma$ ({\it
  solid line}) and $2\sigma$ ({\it dashed line}) contours. The
  parameters are the same as those used in Fig.~\ref{Fig:zsigrlap},
  with the requirement that $r{\rm lap} \ge 5$. 
\label{Fig:ztres_covar}}
\end{figure}

The determination of redshift and age is intrinsically connected,
and in principle one should marginalize over one 
parameter to infer the other. Marginalizing out the redshift (a
continuous variable) to infer the age is straightforward, but the
reverse is more complex, as it involves marginalization over sparsely
sampled variables. The techniques to do this abound in the Bayesian
literature, but we have yet to implement them in \snid. Nevertheless,
we illustrate the covariance between redshift and age using the
2D histogram of age versus redshift residuals, for
correlations satisfying $r{\rm lap} \ge 5$ (Fig.~\ref{Fig:ztres_covar}). As
expected (see \S~\ref{Sect:zreserr}), over(under)-estimating the
age leads to under(over)-estimating the redshift, since the loci of
maximum absorption shift to the red with age (Fig~\ref{Fig:vabst}).

\begin{figure}
\epsscale{1.2}
\plotone{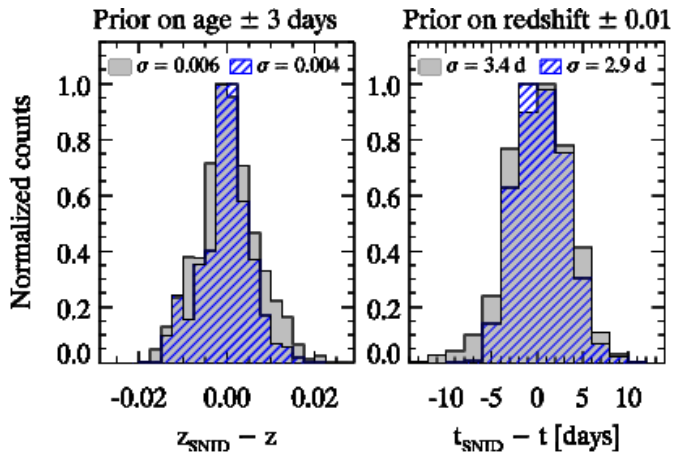}
\caption{Effect of age and redshift constraints on redshift ({\it
  left}) and age ({\it right panel}) residuals, respectively, with the same
  parameters as in Fig.~\ref{Fig:zsigrlap}. Here $z_{\rm \snid}$
  ($t_{\rm \snid}$) corresponds to the median of all redshifts
  (ages) with $r{\rm lap} \ge 5$. The open and hatched histograms
  correspond to residuals with no constraint and a constraint on $(z,t)$,
  respectively. {\it [See the electronic version of the Journal for a
  color version of this figure.]}
\label{Fig:deltazthist_prior}}
\end{figure}

The anti-correlation between redshift and age residuals shown in
Fig.~\ref{Fig:ztres_covar} suggests that constraints on one parameter
should improve the accuracy of the
other. Fig,~\ref{Fig:deltazthist_prior} shows the effect on the
distributions of redshift ({\it left panel}) and age ({\it right panel})
residuals (for $r{\rm lap} \ge 5$; {\it open histograms}) of adding a flat
$\pm 3$-day constraint on age and a flat $\pm 0.01$ constraint on redshift,
respectively ({\it hatched histograms}). A constraint on the
age leads to a $\sim 30\%$ narrower distribution of redshift
residuals (from $\sigma_z = 0.006$ to $\sigma_z = 0.004$) and a
constraint on redshift improves the age determination by $\sim 15\%$
($\sigma_t = 3.4$ days to $\sigma_t = 2.9$ days). In practice, the
constraint on redshift generally comes from a spectrum of the SN host
galaxy and one can impose a constraint on age using a well-sampled
light curve of the supernova (one for which the date of maximum light
is easily determined).

\begin{figure}
\epsscale{1.2}
\plotone{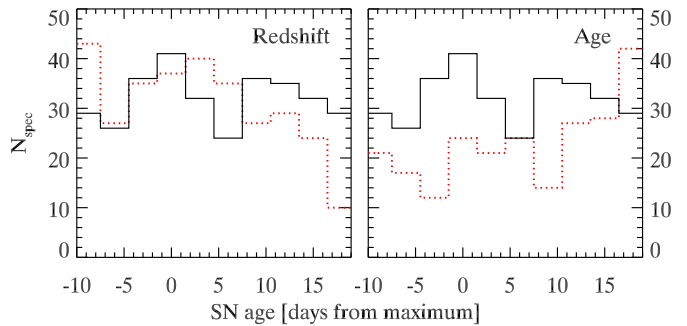}
\caption{Actual ({\it solid line}) and Monte Carlo ({\it dotted
  line}) age distributions of normal SN~Ia templates for redshift
  ({\it left panel}) and age ({\it right panel}) determination. The Monte
  Carlo distribution was computed such that adding more spectra would
  not change the mean redshift and age residuals (for $r{\rm lap} \ge 5$)
  by more than 0.0001 and 0.1 days, respectively.
  {\it [See the electronic version of the Journal for a color version
  of this figure.]}
\label{Fig:opthist}}
\end{figure}

The age distribution of SN spectral templates in the database
affects the accuracy of both cross-correlation age and redshifts.
In Fig.~\ref{Fig:opthist} we show the result of a Monte Carlo simulation
where we compute the number of SN~Ia spectra in bins of 3 days
that would be sufficient for accurate redshift ({\it left panel}) and
age ({\it right panel}) determinations with \snid.
The solid histogram is the actual age distribution of normal SN~Ia
templates in the interval $-10 \le t_B \le +20$ and the dotted
histogram is the Monte Carlo distribution.
We compute 1000 Monte Carlo
realizations for each unity increment in the number of spectra in a
given 3-day age bin. The number of spectra was chosen
such that adding more spectra would not change the mean redshift and
age residuals (for $r{\rm lap} \ge 5$) by more than 0.0001 and 0.1 days,
respectively. For this Monte Carlo distribution, at least eight
correlations with $r{\rm lap} \ge 5$ are needed for the median redshift and
associated error (Fig.~\ref{Fig:deltazhist}) to provide an accurate
estimate.

The Monte Carlo age distribution for redshift determination
(Fig.~\ref{Fig:opthist}, {\it left panel}) has an initial peak around $-10$
days and a bell-shaped envelope roughly centered around maximum ($0$
days)--- akin in fact to a supernova light curve. This is due to the
faster evolution of supernova spectra around maximum light than around
1--2 weeks past maximum (Figs.~\ref{Fig:meanspec}
\&~\ref{Fig:vabst}). In other words, \snid\ can accurately determine
the redshift of an input spectrum at $+10$ days using a template at
$+15$ days, since the wavelength (velocity) positions and relative
strengths of spectral features change little over this age
interval, but will be less accurate when an input spectrum at maximum
light is correlated with template spectra at +5 days, since the
evolution of the spectra is more significant then. The initial peak
around $-10$ days is due to the rapid decrease in spectral line
blueshifts from $\lesssim -10$ days to $\sim -5$ days
(Fig.~\ref{Fig:vabst};
\citealt{Benetti/etal:2005,Blondin/etal:2006a}), rather than to a 
change in the relative strengths of spectral features. 
We are currently lacking normal SN~Ia spectral templates around +5
days past maximum light.
This gap will be filled shortly with a new
set of spectra from the CfA Supernova Program (almost 50 SNe~Ia with
more than 10 epochs of spectroscopy since 2000).

The Monte Carlo age distribution for age determination
(Fig.~\ref{Fig:opthist}, {\it right panel}) is altogether different, but the
same reasons apply: due to the rapid evolution of SN spectra around
maximum light, it is easier to accurately determine the age then
than at 1--2 weeks past maximum, where the spectra evolve on
longer timescales. Hence, more spectra are needed at later ages than
around maximum light. 
The current number of normal SN~Ia templates in our database is
sufficient for accurate age determinations out to $t \lesssim +15$
days, but we need twice the number of templates in the last age
bin. Again, this is within reach with the new set of CfA spectra.

To minimize the impact of our currently non-optimal age distribution
of SN~Ia templates, we have studied the redshift and age residual
distribution when imposing a $1/N_{\rm temp}(t)^a$ weighting scheme,
where $N_{\rm temp}(t)$ is the template age distribution (in 3-day
bins) and $0.0 \le a \le 2.0$ ($a=0.5$ corresponds to a Poisson-like
weighting scheme). This way, the artificial attractors in the actual
age distribution around maximum light and +10 days are down-weighted
with respect to templates at other ages. The weighting scheme does
not lead to any improvement in either the redshift or age
determinations, namely the distribution of residuals for $r{\rm lap} \ge 5$
does not get any narrower. Clearly, a more elaborate method
is necessary to break the redshift-age degeneracy and the
covariance between the two quantities will be included explicitly in a
future version of \snid.

Fig.~\ref{Fig:opthist} only shows the age distribution of
normal SN~Ia templates, for which we have sufficient spectra in
the database to construct a viable Monte Carlo simulation. The faster
evolution of supernova spectra around maximum light is common to all
supernova types, but the homogeneity at a given age may vary
significantly. We are not in a position to test this thoroughly, due
to the limited number of Type Ib/c and Type II templates in the
current \snid\ database. Again, we are confident that new data from
the CfA Supernova Program will better constrain the variance of SN
spectra at a given age (almost 20 SNe~Ib/Ic/II with more than 10
epochs of spectroscopy since 2000). Therefore, while the overall shape
of the Monte Carlo distributions should remain the same, the
absolute scale should be different for the various supernova types.

\subsection{Variation of Redshift and Age Accuracy with Redshift, Age,
  S/N and Galaxy Contamination\label{Sect:zts2ngf}}

The previous studies are valid for the following
parameter space: $0.3 \le z \le 0.5$; $-5 \le t_B \le
+15$; $2 \le$ S/N (per 2\,\AA)$\le 10$. However, we expect the
accuracy of cross-correlation redshifts and ages to change with
redshift, age and signal-to-noise ratio of the input spectrum.

\begin{figure*}
\plottwo{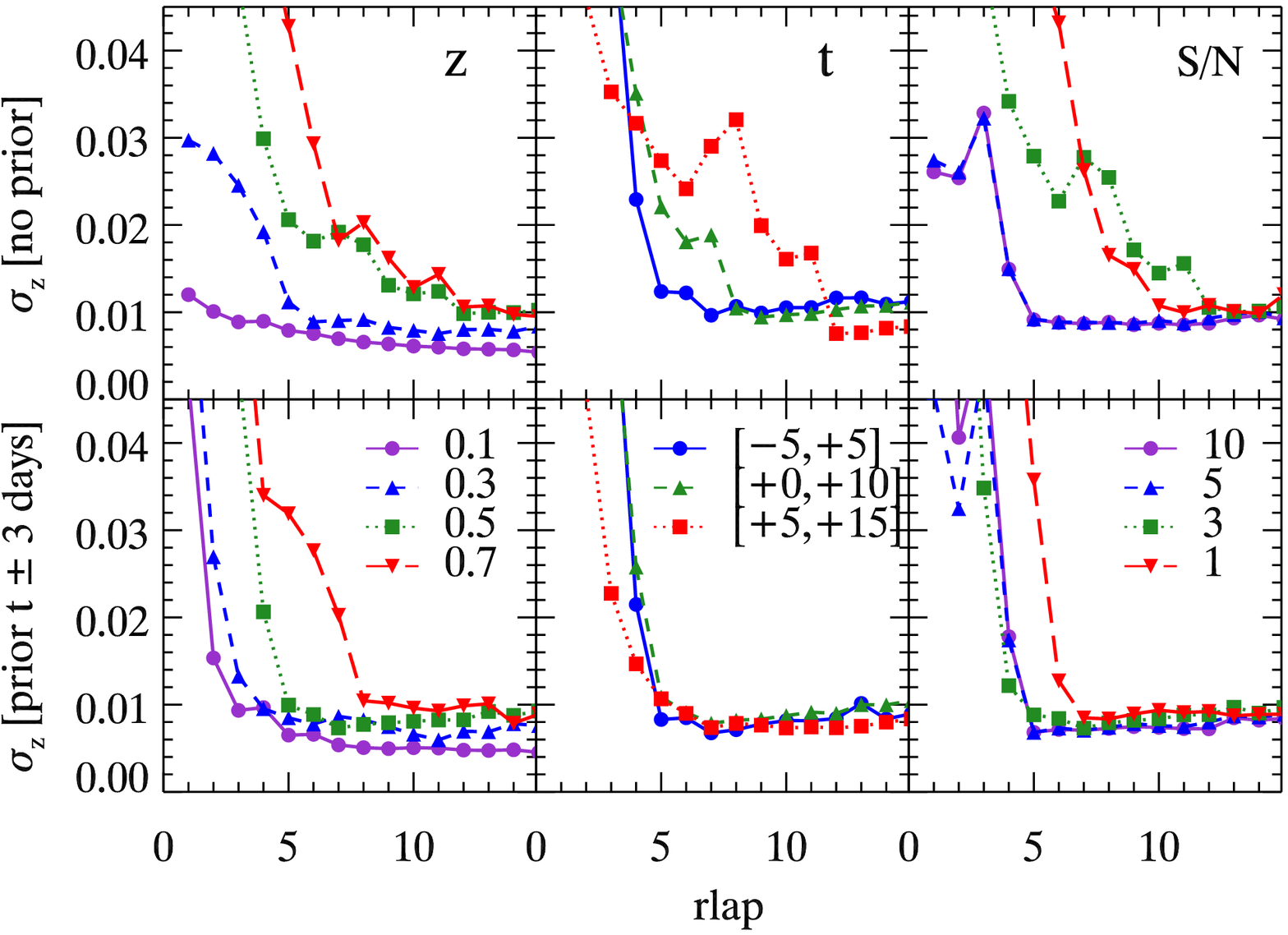}{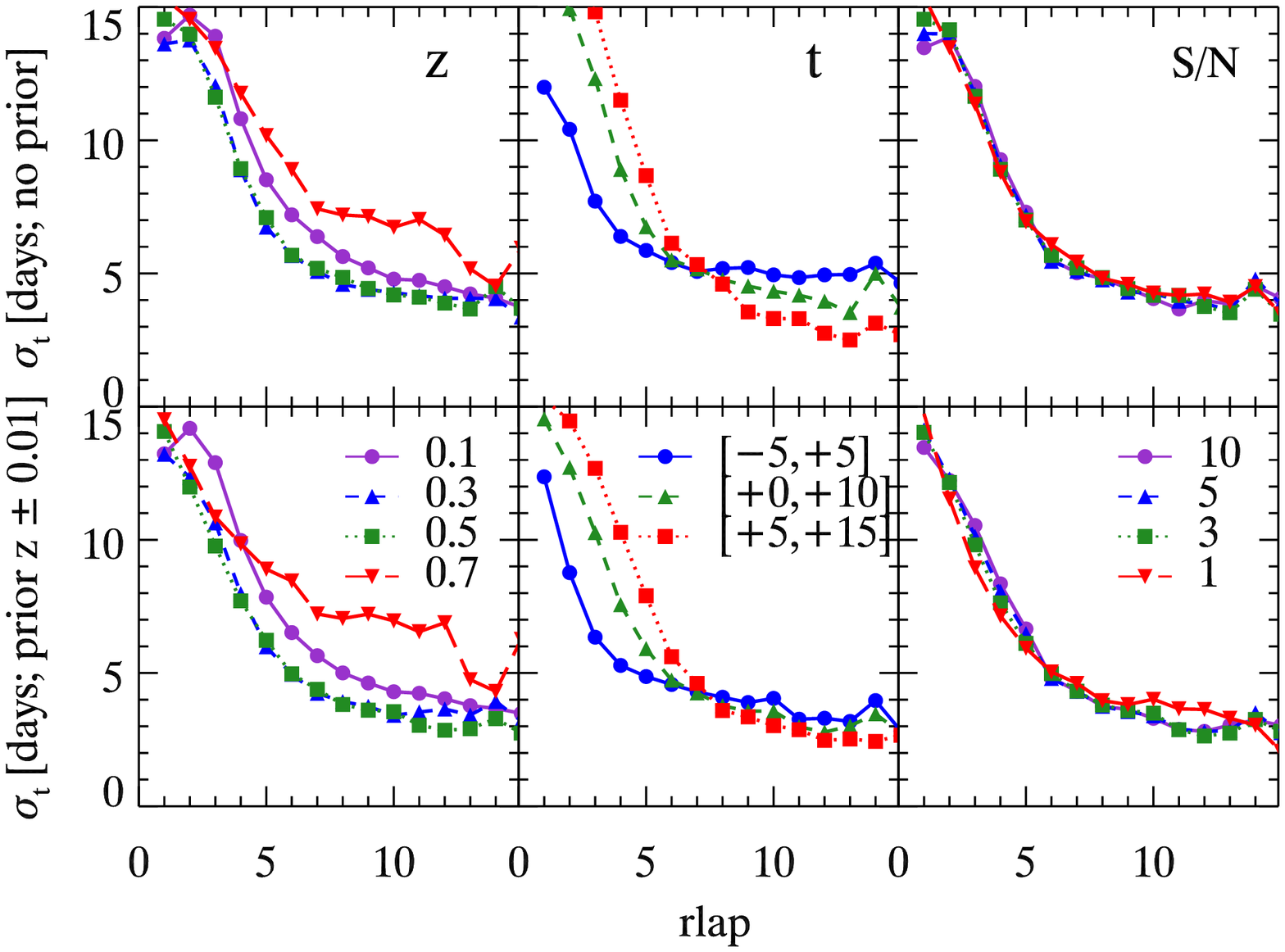}
\caption{{\it Left:} Variation of $\sigma_z$ with redshift ({\it top
  left}), age (in days; {\it top middle}) and S/N (per 2\,\AA; {\it
  top right}). We show the effect of applying 
  a $\pm 3$ day constraint on age in the bottom panels. {\it Right:} Same
  as the left group of panels, but for the variation of $\sigma_t$ with redshift ({\it
  top left}), age ({\it top middle}) and S/N
  ({\it top right}). We show the effect of applying a $\pm 0.01$
  constraint on redshift in the bottom panels. {\it [See the electronic
  version of the Journal for a color version of this figure.]}
\label{Fig:ztvar_zts2n}}
\end{figure*}

In top panels of the left group of panels of  Figure~\ref{Fig:ztvar_zts2n} we show the
variation of the standard deviation of redshift residuals, $\sigma_z$,
with the $r{\rm lap}$ quality parameter for varying redshift ({\it left}), age ({\it middle}),
and S/N ({\it right}). We expect the degrading accuracy with redshift,
since the rest-frame overlap (${\rm lap}$) between the input and 
template spectra in our database decreases with redshift. Even
requiring that ${\rm lap} \ge 0.4$ can lead to degenerate redshifts at the
higher end of the redshift range ($z \gtrsim 0.5$). Increasing the
number of spectra extending blueward to $\l \gtrsim 2000$\,\AA\ would
partially alleviate this problem, although the flux is strongly
depleted at these wavelengths (due to line-blanketing from iron-group
elements) and the most prominent features in supernova spectra are at
optical wavelengths. UV spectra of (nearby) supernovae are rare, but
the database could be expanded at these wavelengths (for SNe~Ia, at
least) by including the higher-S/N publicly-available 
spectra of ongoing high-$z$ SN searches, such as the spectra from the
ESSENCE project \citep{Matheson/etal:2005}.

The variation of $\sigma_z(r{\rm lap})$ with S/N of the
input spectrum (Fig.~\ref{Fig:ztvar_zts2n}, {\it left group of panels;
  top right panel}) is also expected, with a significant degradation below S/N
$\lesssim 3$ per 2\,\AA. The degradation with increasing age of the
input spectrum is again due to the slower evolution of SN spectra at
later ages. The input spectrum will correlate well with template
spectra over a larger range of ages, where  the scatter in the
velocity location of spectral features will translate directly into an
error in redshift.

In the bottom panels of the left group of panels of
Figure~\ref{Fig:ztvar_zts2n} we show 
the effect of applying a flat $\pm 3$ day age constraint on the
$\sigma_z(r{\rm lap})$ curves. The improvement is significant in all cases
(although less so for $z=0.7$).

In the top panels of the right group of panels
Figure~\ref{Fig:ztvar_zts2n} we show the 
variation of the standard deviation of age residuals, $\sigma_t$,
with the $r{\rm lap}$ quality parameter for varying redshift ({\it left}), age ({\it middle}),
and S/N ({\it right}). Again, the degradation at the highest redshift
($z=0.7$) is expected, although it is surprising that the
$\sigma_t(r{\rm lap})$ curves for $z=0.3$ and $z=0.5$ lie atop the one
corresponding to $z=0.1$. It appears that the optimal rest frame
wavelength range of the input spectrum is different for redshift and
age determination. This has already been mentioned by
\citet{Foley/etal:2005} concerning the age determination and
points towards the need for an age- and wavelength-dependent
${\rm lap}(t,\l)$ parameter to weight the correlation height-noise ratio, $r$, instead of
the constant ${\rm lap}$ currently implemented in \snid. The difficulty of
determining the age of an input spectrum at later times is due to
the less rapid evolution of the spectra at these ages. At high
values of $r{\rm lap}$ ($\gtrsim 7$), however, $\sigma_t$ {\it decreases}
with age. This behavior is unexpected, given the discussion in
\S~\ref{Sect:ztcovar} and could again be due to the
wavelength-independent nature of our ${\rm lap}$ parameter.

Even more surprising is the apparent independence of $\sigma_t(r{\rm lap})$
on S/N: for fixed redshift and age (here $z=0.5$
and $-5 \le t_B \le +5$), the $r{\rm lap}$ quality parameter gives an {\it absolute}
measure of the age accuracy, regardless of the S/N of the input
spectrum. Of course, the probability of having correlations with high
$r{\rm lap}$ values drops with S/N, but we have checked that our simulation
yielded a sufficient number of correlations at $r{\rm lap} \gtrsim 7$ for
this result to be statistically significant.

We next study the impact of contamination from the underlying spectrum
of the host galaxy affecting the input supernova spectrum. The
contamination fraction will depend both on the projected position of
the supernova within its host (higher contamination closer to the
nucleus) and on the relative flux difference between the supernova and
the portion of the galaxy located in the same aperture
(i.e. immediately underlying the SN trace, when extracting the
spectrum). Several techniques are commonly used to separate the
supernova light from that of the host galaxy, either through galaxy
template subtraction (e.g., in the algorithm presented by
\citealt{Howell/etal:2005}), or using more elaborate techniques such
as two-channel deconvolution directly applied to the 2D spectrum
\citep{Blondin/etal:2005}. However, neither of these techniques works
well in cases where the SN lies on top of the nucleus of a bright
galaxy and in all other cases there still remains some fraction of
galaxy light in the SN spectrum.

\begin{figure}
\epsscale{1.1}
\plotone{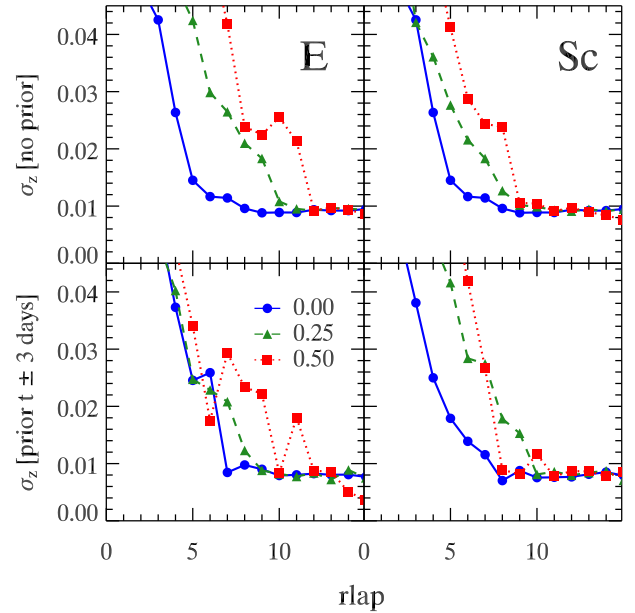}
\caption{{\it Top:} Impact of galaxy contamination fraction on
  $\sigma_z(r{\rm lap})$, for both elliptical (E; {\it left panel}) and spiral
  (Sc; {\it right panel}) galaxies. The different lines correspond to
  different galaxy contamination fractions, expressed in fractions of
  the total flux. {\it Bottom:} Effect of applying a $\pm 3$-day
  constraint on the age. {\it [See the electronic version of the
  Journal for a color version of this figure.]} 
\label{Fig:zvar_gf}}
\end{figure}

We contaminate each input spectrum in our simulation with galaxy
light, using the elliptical and Sc galaxy templates of
\citet{Kinney/etal:1996}. The top panels of Figure~\ref{Fig:zvar_gf} 
 show the effect on redshift residuals,
$\sigma_z(r{\rm lap})$, of increasing the galaxy contamination fraction
(expressed in fractions of the total flux) from 0.00 to 0.50. The
impact of the elliptical galaxy ({\it top left panel}) is most
severe, since the spectra of early-type hosts contain broad continuum
structures that yield strong power at similar wavenumbers as
supernova features in Fourier space. Late-type galaxies have smoother
continua and their narrow emission lines are filtered out using the
bandpass filter. In the bottom panels of Fig.~\ref{Fig:zvar_gf} we
apply a flat age constraint of $\pm 3$ days. The improvement, if
any, is negligible for both the elliptical and Sc galaxy types.

We have run several simulations to test whether the $r{\rm lap}$ quality parameter
could be used to evaluate the amount of galaxy contamination for
various galaxy types, but the results were inconclusive. This
constitutes the real limit of \snid:
some extra pre-processing of the supernova spectrum is
necessary to ensure that the input to \snid\ is as ``clean'' as
possible. Other algorithms (see \S~\ref{Sect:comparison}) perform a
simultaneous fit of the galaxy fraction when comparing the input SN
spectrum to the set of templates in the database, which enables
the classification of supernovae when the galaxy contamination is
$\lesssim 75\%$ \citep{Howell/etal:2005}.

\subsection{Comparison with External Measurements}

In this section we test the accuracy of correlation redshifts using
\snid\ by comparing them with those of the host galaxy. Galaxy redshifts
($z_{\rm GAL}$) are routinely determined using nebular emission lines
in their spectra or by cross-correlation with absorption-line galaxy
spectral templates \citep{Kurtz/Mink:1998}. They are typically
accurate to $< 0.001$. However, the redshift of the supernova can
differ from $z_{\rm GAL}$, since the supernova event may have
occurred in a region where its velocity (in the galaxy rest frame) is
different from the mean value, due to the velocity dispersion of the
galaxy's light-emitting component ($\sim 100$\kms\ and
$\sim 200$\kms\ in early- and late-type galaxies, respectively;
\citealt{McElroy:1995}). Nevertheless, $z_{\rm GAL}$ gives
a more accurate determination of the SN redshift than \snid\ (which
has typical redshift errors of $\lesssim 0.01$ for $r{\rm lap} \ge 5$), so a
comparison of the two gives a valuable indication on the accuracy of
\snid\ redshifts, determined from real data.

\begin{figure*}
\begin{center}	
\plottwo{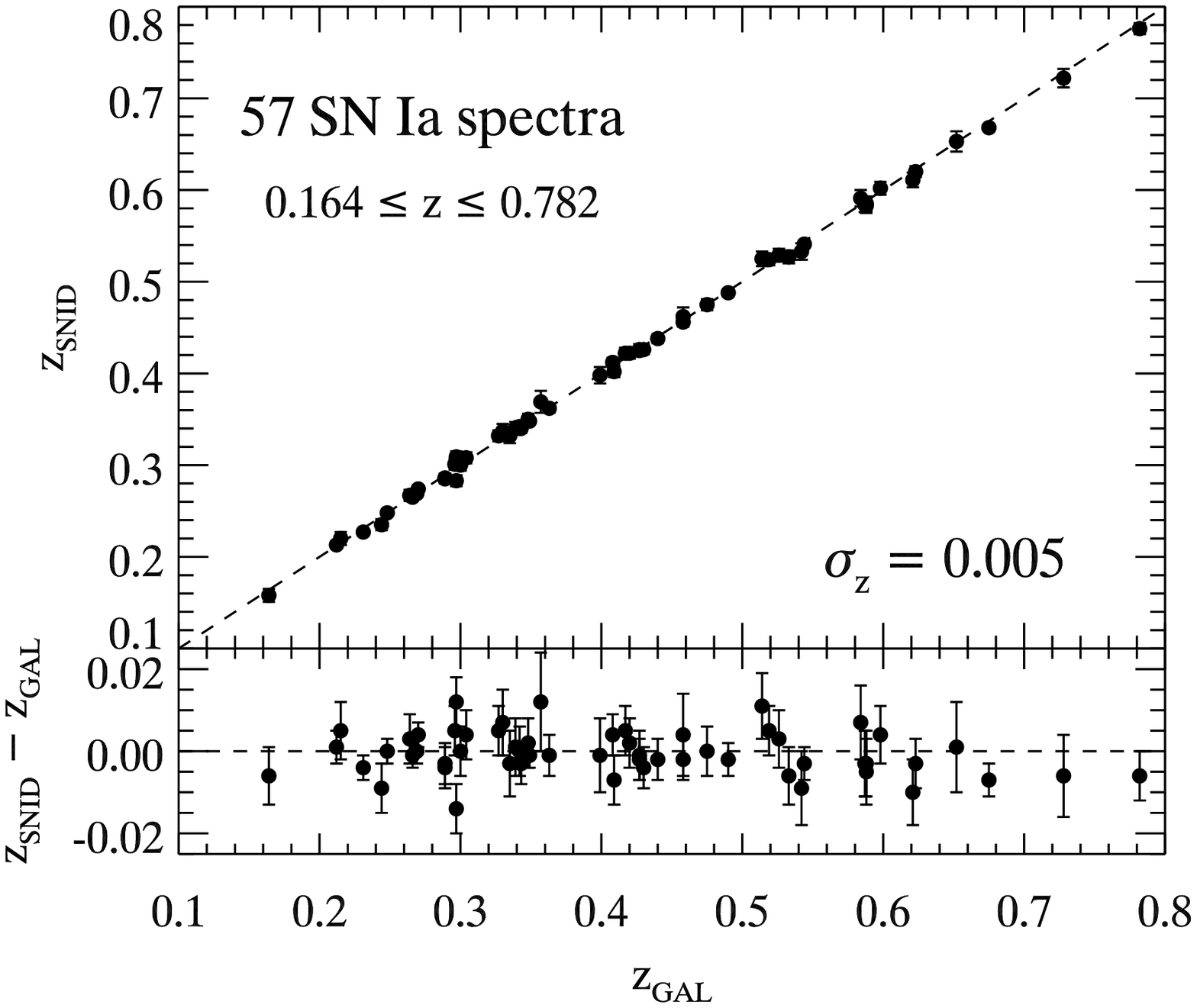}{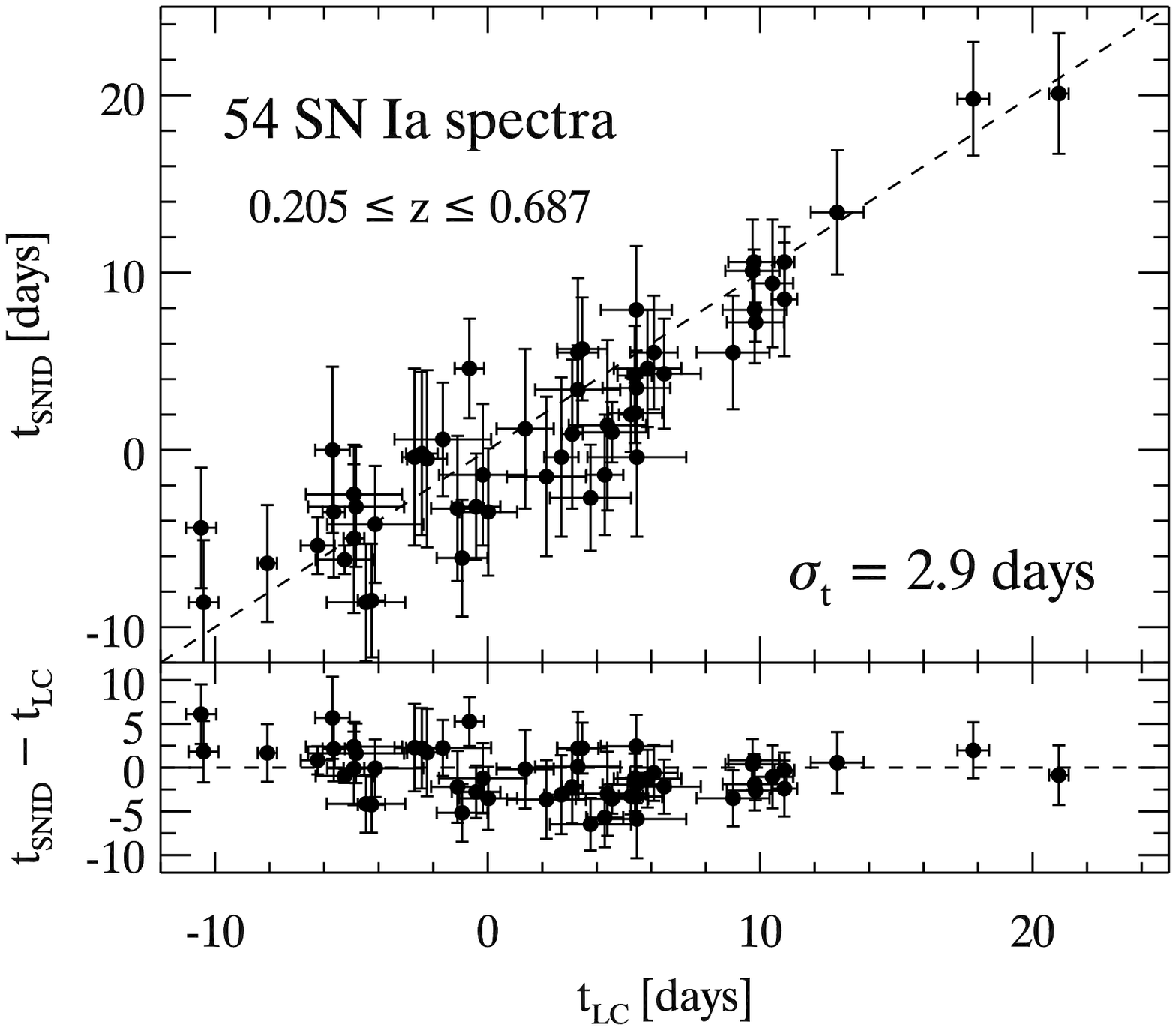}
\caption{{\it Left:} Comparison of redshifts determined from
  cross-correlations with SN~Ia spectral templates using \snid\
  ($z_{\rm \snid}$) and from narrow lines in the
  host galaxy spectrum ($z_{\rm GAL}$; {\it top}). We show the
  residuals vs. $z_{\rm GAL}$ in the bottom panel. {\it Right:}
  Comparison of supernova spectral ages determined using \snid\
  ($t_{\rm \snid}$) and rest-frame light-curve ages ($t_{\rm LC}$)
  of high-$z$ SNe~Ia ($0.164 \le z \le 0.587$; {\it top}). We
  show the residuals vs. $t_{\rm LC}$ in the bottom panel. The
  data are from the ESSENCE project
  (\citealt{Matheson/etal:2005,Miknaitis/etal:2007}; Foley et al.,
  {\it in prep}; Blondin et al., {\it in prep}). 
\label{Fig:zgaltlcsnid}}
\end{center}
\end{figure*}

We have selected high-redshift SN~Ia spectra taken by the
ESSENCE team (\citealt{Matheson/etal:2005}; Foley et al., {\it in
  prep}; Blondin  et al., {\it in prep}), for which a redshift of the
host galaxy was obtained. This amounts to 57 SN~Ia spectra in
the redshift range $0.164 \le z \le 0.782$. The result of this
comparison is shown in the left panels of Figure~\ref{Fig:zgaltlcsnid}. The
dispersion about the one-to-one correspondence of the redshifts is
excellent, with $\sigma_z \approx 0.005$ over the whole redshift
range. This is in good agreement with the expected redshift residual
found from simulations (with no constraint on the age;
Fig.~\ref{Fig:deltazthist_prior}). The bottom left panel shows a plot of the
redshift residuals as a function of the galaxy redshift. The mean
residual is $\sim 4 \times 10^{-4} \ll \sigma_z$, which shows that
there are no systematic effects in using \snid\ to determine the SN
redshift.

To compare the supernova age determined through cross-correlation
with external measurements, we select ESSENCE high-redshift SN~Ia
spectra for which a well-sampled light curve is available around maximum
light \citep{Miknaitis/etal:2007}. This way we can determine the time
difference (in the {\it observer} frame) between maximum light,
($t_{\rm max}$) and the time the spectrum was obtained ($t_{\rm spec}$)
and compare this time interval with the {\it rest-frame} age ($t_{\rm
\snid}$) determined through cross-correlation with local SN~Ia
templates. For this comparison to make sense we must correct the
light-curve age for the $(1+z)$ time-dilation factor expected in an
expanding universe
\citep{Wilson:1939,Rust:1974,Leibundgut/etal:1996,Goldhaber/etal:2001}.
We expect a one-to-one correspondence between

\begin{equation}
t_{\rm LC} = \frac{t_{\rm spec}-t_{\rm max}}{1+z}
\end{equation}

\noindent
and $t_{\rm \snid}$. The result is shown in the right panels of
Figure~\ref{Fig:zgaltlcsnid}. We used a total of 54 spectra in the
redshift range 
$0.205 \le z \le 0.687$, 27 of which had an associated galaxy
redshift--- which we used as a constraint when determining
the age. The dispersion about the $t_{\rm \snid} = t_{\rm LC}$
line is $\sigma_t \approx 2.9$ days over a age interval $-10 \lesssim
t_{\rm LC} \lesssim +20$, again in good agreement with the expected
residuals (Fig.~\ref{Fig:deltazthist_prior}). We show the
residuals versus $t_{\rm LC}$ in the bottom right panel. The mean  residual
is approximately $-0.7$ days. 
The excellent correspondence between $t_{\rm
  LC}$ and $t_{\rm \snid}$ shows that \snid\ can be used in studies of
time dilation effects in high-redshift multiepoch SN~Ia spectra 
\citep{Riess/etal:1997,Foley/etal:2005}.

The correlation technique could not have yielded such good results had
the high-$z$ SNe~Ia in the sample been significantly different from
the SN~Ia template spectra in the \snid\ database. The fact that the
correlation redshifts and ages agree so well with the galaxy
redshifts and light-curve ages, respectively, is a strong argument
in favor of the similarity of 
these SNe~Ia with local counterparts.


\section{Type Determination}\label{Sect:type}

The results of \S~\ref{Sect:zt} are only valid if we assume that we know
the type of the input supernova spectrum--- in this case a normal
SN~Ia. Although \snid\ is tuned to determining SN redshifts, we
investigate its potential in determining the SN type in an impartial
way. We base our investigation on a simple frequentist
approach as opposed to a more elaborate Bayesian one, but we discuss the
future implementation of the latter in \snid\ in
\S~\ref{Sect:comparison}.

In what follows we focus on five distinct examples, the first three
being particularly relevant to ongoing high-redshift SN~Ia searches:
the distinction between 1991T-like SNe~Ia and other SNe~Ia
(\S~\ref{Sect:1anorm1a91t}); the distinction  
between SNe~Ib/c and SNe~Ia at high redshifts
(\S~\ref{Sect:1cnorm1a}); the identification of peculiar SNe~Ia
(\S~\ref{Sect:1apec}); finally, the distinction between SNe~Ib and
SNe~Ic and between SNe~IIb and both SNe~II 
and Ib (\S~\ref{Sect:otheregs}), more relevant to ongoing nearby ($z
\lesssim 0.1$) supernova searches. We used the same simulation setup
as in \S~\ref{Sect:simulation}, except we consider correlations with
all supernova types in the database.

The reader must keep in mind that, while the age distribution of
SN~Ia templates is close to optimal in the current \snid\ database
(see \S~\ref{Sect:ztcovar}), those for supernovae of other types are
most certainly not. While the results presented in this section are
encouraging, they are no doubt biased by the relatively low number of
SN~Ib/Ic/II templates with respect to SNe~Ia.

\subsection{Normal versus 1991T-like SNe Ia\label{Sect:1anorm1a91t}}

It can be challenging to distinguish the subtypes of SNe~Ia from one
another. 1991T-like SNe~Ia have a peak luminosity at the bright end of
the SN~Ia distribution and although their light curves still obey the
width-peak luminosity, or ``Phillips,'' relation
\citep{Phillips:1993}, it is useful to have an independent  
confirmation of their high intrinsic luminosity from their
spectra. Spectra of 1991T-like SNe~Ia are characterized by the
near-absence of \catwo\ and \sitwo\ lines in the early-time spectra,
and prominent high-excitation features of \fethree--- not found in
normal SNe~Ia. The \sitwo, \stwo and \catwo\ features develop
during the post-maximum ages and by $\sim 2$ weeks past maximum the
spectra of ``1991T-like'' objects are similar to those of normal
SNe~Ia
\citep{Filippenko/etal:1992,Ruiz-Lapuente/etal:1992,Phillips/etal:1992,Jeffery/etal:1992}.

\begin{figure*}
\begin{center}	
\includegraphics[width=7in]{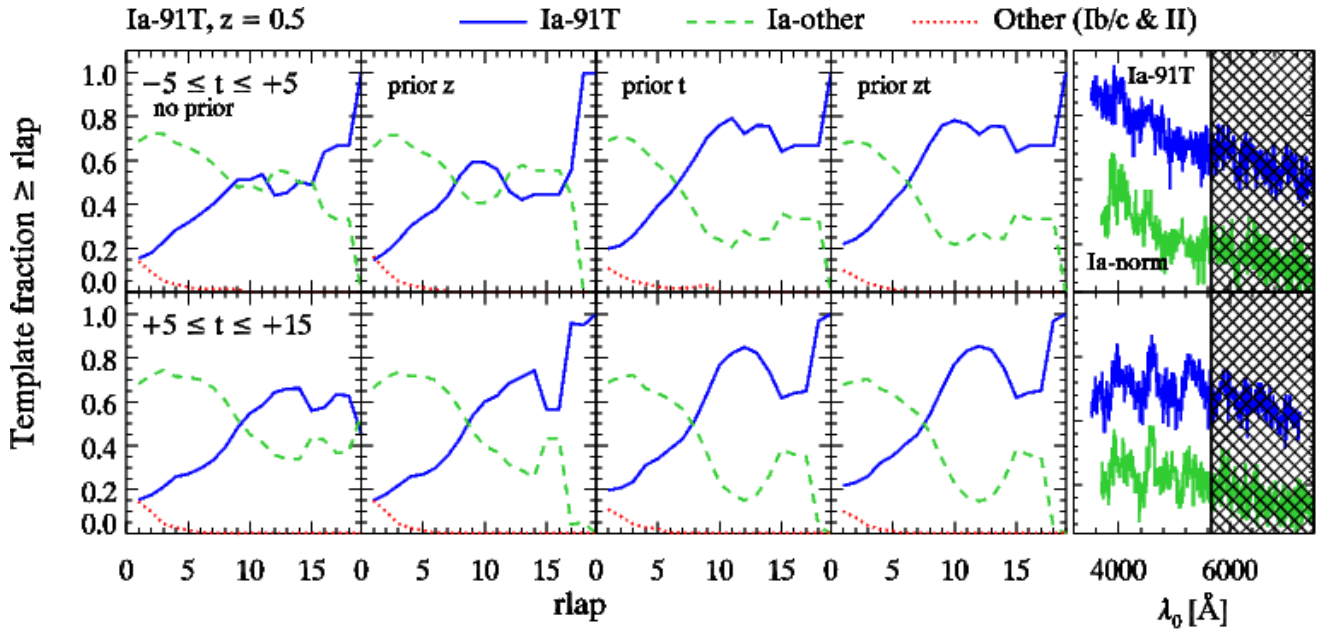}
\caption{Attempt to identify a 1991T-like SN~Ia at $z=0.5$ in the
  age interval $-5 \le t_B \le +5$. {\it Top:}
  Fraction of templates in the \snid\ database corresponding to a
  certain supernova type (1991T-like SN~Ia: {\it solid line}; SN~Ia of
  other subtypes: {\it dashed line}; supernova of other types: {\it
  dotted line}), in $r{\rm lap}$ bins of size unity. {\it Left to
  right:} With no constraints on the redshift or age, with a $\pm
  0.01$ constraint on the redshift, with a $\pm 3$ days constraint on
  the age, and with a combined constraint on redshift and age.
  {\it Bottom:} Same lines as above, but for post-maximum
  spectra in the age interval $+5 \le t_B \le +15$.
  The right panel shows representative spectra of 1991T-like and
  normal SNe~Ia, around maximum light ({\it top}) and
  $\sim$1--2 weeks past maximum ({\it bottom}), as observed with a
  typical optical spectrograph at $z=0.5$ (the cross-hatched area represents
  the rest frame portion of the spectrum that is lost due to the
  redshift). Note that the relative differences in {\it
  pseudo}-continuum shapes have no impact on the \snid\ results. {\it
  [See the electronic version of the Journal for a color version of
  this figure.]} 
\label{Fig:Ia91t}}
\end{center}
\end{figure*}

In the top panels of Figure~\ref{Fig:Ia91t} we illustrate the ability
of \snid\ to identify 1991T-like SNe~Ia at $z=0.5$. The input
spectrum is a ``Ia-91T'' template in the \snid\ database 
(Table~\ref{Table:sndb}) in the age interval $-5 \le t_B
\le +5$, i.e. when the spectroscopic differences with normal
SNe~Ia are most apparent. We show the fraction of templates in the
\snid\ database that correlate with the input spectrum, as a function
of the $r{\rm lap}$ quality parameter: 1991T-like SNe~Ia ({\it solid line}),
other SNe~Ia ({\it dashed line}), and supernovae of other types
({\it dotted line}). From left to right, we show the effect of
having no constraint on either age or redshift, a flat $\pm 0.01$ constraint
on redshift, a flat $\pm 3$ day constraint on the age, and a combined constraint
on both redshift and age.

When there is no constraint on either redshift or age, the fraction of
Ia-91T templates ({\it solid line}) for $r{\rm lap} \gtrsim 10$ is
greater than that for other SN~Ia templates ({\it dashed line}). Note
that for $r{\rm lap} \gtrsim 5$ the confusion 
with supernovae of other types is practically non-existent ($\lesssim
2\%$). Adding a constraint on the redshift does not lead to a
  significant improvement, but a constraint on age reduces the cross-over
 $r{\rm lap}$ value between other SN~Ia and 1991T-like templates from
  $r{\rm lap} \gtrsim 15$ to $r{\rm lap} \approx 6$.
The negative noise spikes around $r{\rm lap} \approx 15$ are
statistical noise due to the small number of templates with $r{\rm lap}$
values in excess of that cutoff.

The bottom panels of Figure~\ref{Fig:Ia91t} show the same lines for
input spectra in the age interval $+5 \le t_B \le +15$. At
these post-maximum ages, the differences between 1991T-like and
normal SNe~Ia are less apparent (see the rightmost panel) and the
impact on the ability of \snid\ to distinguish between the different Ia
subtypes is readily apparent. In the absence of constraints on redshift or
age, the fraction of Ia-91T templates reaches a peak of 
  $\sim 70$\%, while it increases to 100\% with constraints
on redshift and age. The variation for $r{\rm lap} \ge 15$
is again statistical noise due to the limited number of templates with
such good correlations.

The difficulty of distinguishing between normal and 1991T-like
SNe~Ia at high redshift could partly explain the apparent lack of
1991T-like SNe~Ia at high redshifts ($2/52 \approx 4$\% in the
SN~Ia sample published by \citealt{Matheson/etal:2005}) with respect
to the fraction expected locally (up to $\sim20$\% according to
\citealt{Li/etal:2001b}).

\subsection{SN Ia versus SN Ib/c\label{Sect:1cnorm1a}}

The misidentification of supernovae of other types as SNe~Ia is a
major concern for ongoing high-redshift SN~Ia searches. Including only
a small fraction of non-Ia supernovae in a sample would lead to a
mis-calibration of the absolute magnitudes of these objects and to
biases in the derived cosmological parameters \citep{Homeier:2005}. A
particular concern is the contamination of high-z SN~Ia samples with
SNe~Ib/c. At redshifts $z \gtrsim 0.4$, the defining \sitwo\
$\l 6355$ absorption feature of SNe~Ia (also present, although somewhat
weaker, in SNe~Ic) is redshifted out of the range of most optical
spectrographs and one has to rely on spectral features blueward of
this to determine the supernova type. Some of these features, such as
the \catwo\ H and K $\l \l 3934, 3968$ doublet, are common to both SNe~Ia
and SNe~Ib/c. Other features characteristic of SN~Ia spectra around
maximum light (e.g. \stwo\ $\l \l 5454, 5640$) are generally weak and
can be difficult to detect in low-S/N spectra. One has to invoke
external constraints, such as the SN color evolution, light-curve
shape, host galaxy morphology (only SNe~Ia occur in early-type hosts;
\citealt{Cappellaro/etal:1997}), or the expected apparent peak
magnitude: SNe~Ib/c at maximum light are often $\gtrsim
1$~mag fainter than SNe~Ia (\citealt{Richardson/etal:2006};
although \citealt{Clocchiatti/etal:2000} have reported on one SN~Ic
with an absolute magnitude similar to normal SNe~Ia) and hence are
only expected to ``pollute'' magnitude-limited samples of SNe~Ia at
the lower-redshift end. If the redshift is not known, SNe~Ib/c can be a
serious contaminant for high-redshift SN~Ia searches.

\begin{figure*}
\begin{center}	
\includegraphics[width=7in]{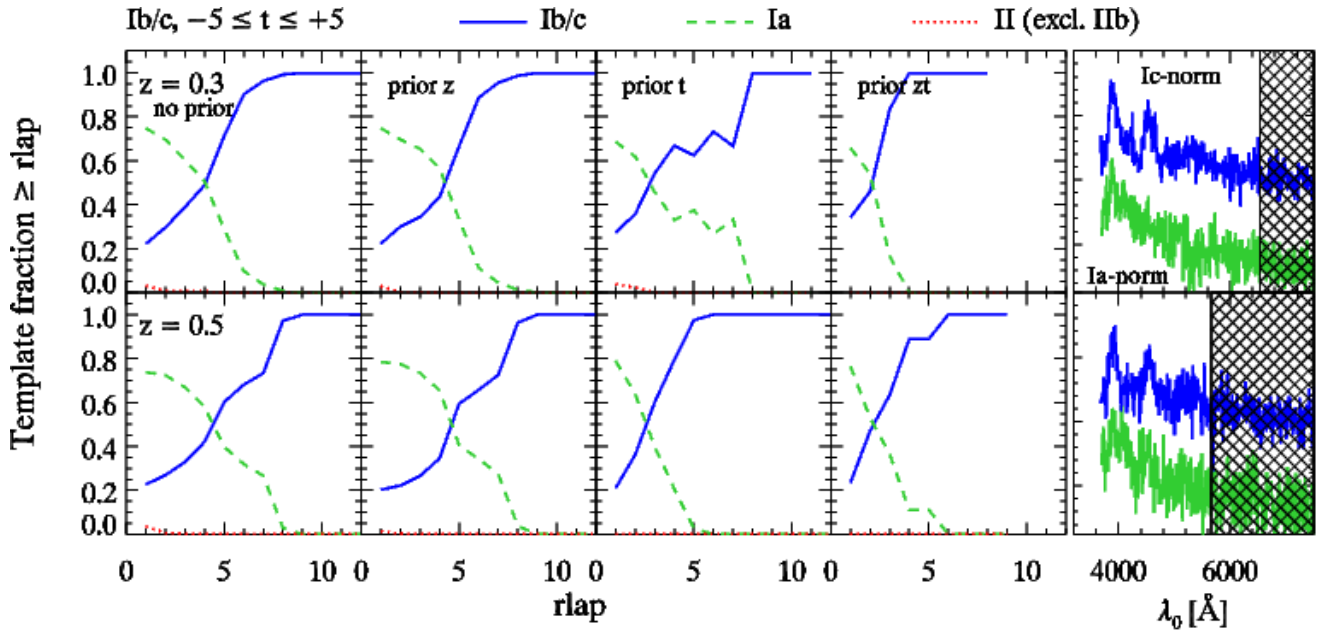}
\caption{Same as Fig.~\ref{Fig:Ia91t}, but for normal SNe~Ic
  in the age interval $-5 \le t_B \le +15$, at
  $z=0.3$ ({\it top}) and $z=0.5$ ({\it bottom}). Here
  the lines correspond to fractions of SNe~Ib/c ({\it solid
  line}), SNe~Ia of all subtypes ({\it dashed line}), and SNe~II
  (excluding SNe~IIb, {\it dashed line}). {\it [See the electronic
  version of the Journal for a color version of this figure.]}
\label{Fig:Icnorm}}
\end{center}
\end{figure*}

In the top panels of Figure~\ref{Fig:Icnorm} we illustrate the ability
for \snid\ to identify SNe~Ib/c at $z=0.3$. The input spectrum is a
``Ib-norm'' or ``Ic-norm'' template in the \snid\ database 
(Table~\ref{Table:sndb}) in the age interval $-5 \le t_B
\le +5$. We show the fraction of templates in the \snid\ database that
correlate with the input spectrum, as a function of the $r{\rm lap}$
quality parameter: all SNe~Ib/c (including SNe~IIb; {\it solid line}), SNe~Ia ({\it dashed
  line}), and SNe~II (excluding SNe~IIb; {\it dotted line}).

In the absence of a constraint on age, correlations with $r{\rm lap}
\gtrsim 4$ are sufficient to recover a dominant fraction of SNe~Ib/c over
SNe~Ia. The confusion with SNe~II is practically nonexistant ($<5\%$
for $r{\rm lap} \lesssim 3$; 0\%\ for $r{\rm lap} > 3$). A constraint on age only
reduced the cross-over $r{\rm lap}$ between SNe~Ia and SNe~Ib/c from
$r{\rm lap} \approx 4$ to $r{\rm lap} \approx 3$ but leads to less correlations
with $r{\rm lap} \gtrsim 5$ (hence the noise spikes in the recovered
template fractions). A combined constraint on redshift and age yields
a 100\%\ SN~Ib/c fraction for $r{\rm lap} \ge 4$, but no correlations with
$r{\rm lap} > 8$.


At $z=0.5$, the results are essentially unchanged from $z=0.3$,
although the absolute number of good correlations ($r{\rm lap} \gtrsim 5$)
is generally less. We also note that these results are biased by the
confusion between SNe~Ib and SNe~Ic at this redshift (up to
$\sim40\%$). The constraint on age leads to a more significant 
improvement than at $z=0.3$, suggesting that the similarity between
spectra of SNe~Ic around maximum light and SNe~Ia at  $\sim$1--2
weeks past maximum is less problematic over this restricted
wavelength range.


Note that the mis-classification of SNe~Ib/c as SNe~Ia (or the reverse)
can sometimes pose problems with the high-S/N spectra of nearby
objects, especially at later ages or if no age information is
available. A striking example is the nearby SN~Ic SN~2004aw
\citep{Taubenberger/etal:2006}, originally classified as an SN~Ia
 by \citet{Benetti/etal:2004b}. More recently, two nearby
supernovae originally announced as Type Ic events around maximum light
(SN~2006bb and SN~2006bk;
\citealt{Kinugasa/Yamaoka:2006a,Kinugasa/Yamaoka:2006b}) were
re-classified as SNe~Ia at 2--3 weeks past maximum light
based on cross-correlations with SN spectra of all types using \snid\
\citep{Blondin/etal:2006b,Blondin/etal:2006c}.

\subsection{Identifying SN~Ia ``Oddballs''\label{Sect:1apec}}

Some SNe~Ia, which we refer to as peculiar, do not
belong to any of the normal, 1991T-like, or 1991bg-like
categories. Such is the case of SNe~2000cx \citep{Li/etal:2001a}, 2002cx
\citep{Li/etal:2003} and 2005hk \citep{Phillips/etal:2007}. The first
of these has pre-maximum spectra similar to those of SN~1991T,
although the \sitwo\ lines that appear around maximum light remain
strong several weeks past maximum. SN~2002cx, 
``the most peculiar known SN~Ia''
\citep{Li/etal:2003,Branch/etal:2004,Jha/etal:2006b} and SN~2005hk are
even more difficult to accommodate in the current classification
scheme: their early-time spectra show signatures of high-ionization
lines of iron, as in the overluminous SN~1991T, but their luminosity
is similar to that of the subluminous SN~1991bg. Moreover, their
$I$-band light curves are devoid of the secondary maximum present in
all other Ia subtypes. Both objects possibly originate from a pure
deflagration explosion (see \citealt{Phillips/etal:2007}) and could
form an altogether separate class of SNe~Ia.

These peculiar SNe~Ia do not obey the Phillips relation and thus
cannot be used as calibrated distance indicators. They must be weeded
out of high-redshift SN~Ia samples in order to avoid significant
biases in the derived cosmological parameters. Until recently, there
was no evidence for such peculiar events at high redshifts. This has
changed with the recent discovery of the overluminous SNLS-03D3bb
(SN~2003fg) at $z=0.244$ \citep{Howell/etal:2006}.

\begin{figure*}
\begin{center}	
\includegraphics[width=7in]{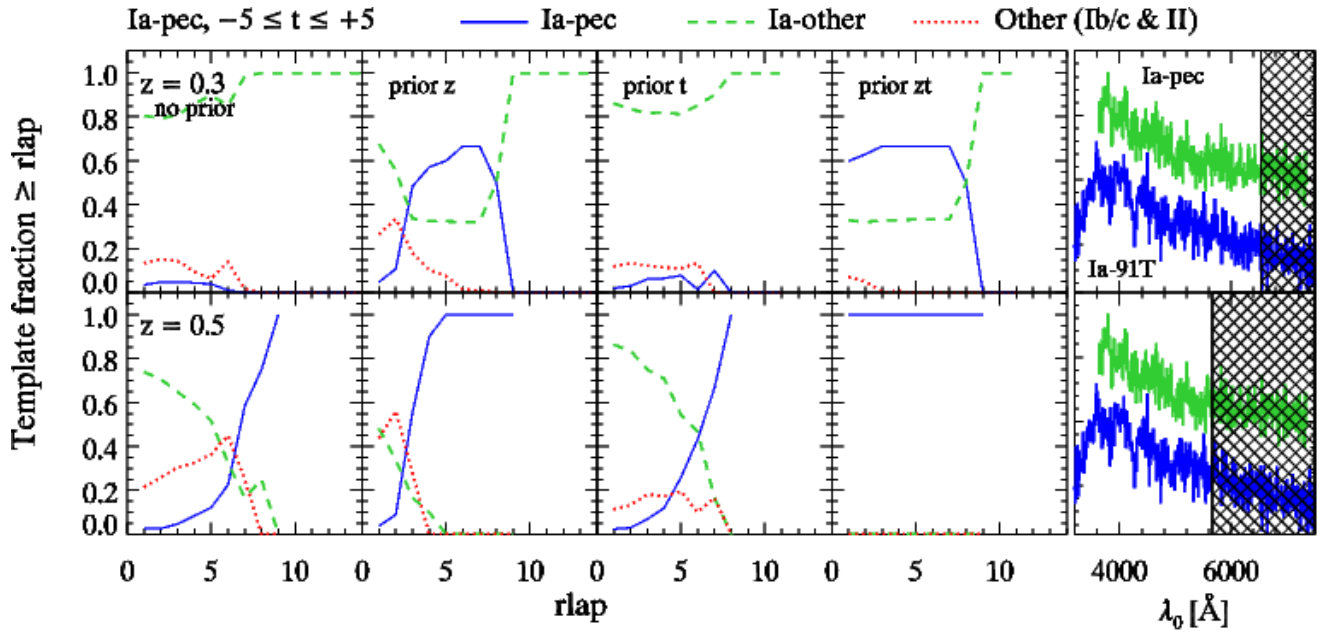}
\caption{Same as Fig.~\ref{Fig:Ia91t}, but for a peculiar SN~Ia
  in the age interval $-5 \le t_B \le +15$, at
  $z=0.3$ ({\it top}) and $z=0.5$ ({\it bottom}). Here
  the lines correspond to fractions of peculiar SNe~Ia ({\it solid
  line}), other SNe~Ia ({\it dashed line}), and all other
  supernova types ({\it dashed line}). {\it [See the electronic
  version of the Journal for a color version of this figure.]} 
\label{Fig:Iapec}}
\end{center}
\end{figure*}

In the top panels of Figure~\ref{Fig:Iapec} we illustrate the ability
of \snid\ to identify peculiar SNe~Ia at $z=0.3$. The input
spectrum is a ``Ia-pec'' template in the \snid\ database 
(Table~\ref{Table:sndb}) in the age interval $-5 \le t_B
\le +5$. We show the fraction of templates in the \snid\ database
that correlate with the input spectrum, as a function of the $r{\rm lap}$
quality parameter: peculiar SNe~Ia ({\it solid line}), other SNe~Ia
({\it dashed line}), and supernovae of other types ({\it dotted line}). In
the absence of a constraint on redshift, the maximum recovered fraction of
Ia-pec templates is $\sim 5$\%, with no correlations at $r{\rm lap} >
6$. With a constraint on redshift, the Ia-pec template fraction peaks
at $\sim 70$\%, but the best correlations are always for another Ia
subtype (most frequently 1991T-like). At $z=0.5$, the recovered
fraction of Ia-pec templates is complete for $r{\rm lap} \ge 5$ for both
a constraint on redshift and a combined constraint on redshift and
age. This apparent improvement is counterbalanced by the absence of
correlations with $r{\rm lap} > 9$. Note that in the absence of any
constraint, the recovered fraction of Ia-pec templates is dominant
for $r{\rm lap} \gtrsim 6$.

Despite the limited number of peculiar SNe~Ia in our database
(SN~2000cx, SN~2002cx and SN~2005hk, for a total of 15 spectra in the
age interval $-5 \le t_B \le +5$; see Table~\ref{Table:sndb}), the
change in template fraction from $z=0.3$ to $z=0.5$ gives indications
as to which rest frame wavelength range is most valuable for
determining the SN type. This again calls for a wavelength (and age)
weighting of the spectrum overlap parameter, ${\rm lap}(\l,t)$ (see \S~\ref{Sect:zts2ngf}), which we
are working to implement in a future version of \snid.

We note that \snid\ unambiguously confirms the similarity between
SN~2005hk and SN~2002cx (see \citealt{Phillips/etal:2007}): for all the
spectra of SN~2005hk in the \snid\ database (Table~\ref{Table:sndb}), the
best-match template spectrum is SN~2002cx, whether a constraint on redshift
and age is applied or not. In the absence of a constraint on redshift,
however, the fraction of 1991T-like templates that correlate with the
input SN~2005hk spectrum increases dramatically, leading to an
overestimation of the redshift by $\sim 0.02$, roughly corresponding
to the difference in absorption velocities between SN~2005hk and
SN~1991T at a given age ($\sim 6000$\kms;
\citealt{Phillips/etal:2007}).

SNLS-03D3bb (SN~2003fg) at $z=0.244$ \citep{Howell/etal:2006}, on
the other hand, illustrates the limitations of the cross-correlation
approach in determining the SN type, when applied to objects that are
not part of the library of spectral templates. Its spectrum (at +2
days; A.~Howell, private communication) is unique among supernova
spectra and we have no similar examples in our database. 
In the absence of constraints, the best-match template is the
1991T-like SN~1999dq \citep{Matheson/etal:2007} at $t=+6.2$ days
($z=0.254 \pm 0.005$). With constraints on age or redshift, the
best-match template is in all cases the normal SN~Ia SN~1989B at
$t=+3.5$ days ($z=0.251 \pm 0.005$). The spectrum of SNLS-03D3bb is
now part of the \snid\ database and will prove useful to identify
such peculiar objects at all redshifts.

\subsection{Further Specific Examples\label{Sect:otheregs}}

We next focus on two further specific examples, relevant to the
spectroscopic classification of supernovae in nearby ($z \lesssim
0.1$) SN searches: the distinction between SN~Ib and Ic,
and that between SN~IIb and II/Ib.

\subsubsection{SN~Ib versus SN~Ic}

SN~Ib and Ic are often difficult to tell apart and are
sometimes referred to as ``SNe~Ib/c'' in the literature
(e.g., SN~1999ex; \citealt{Hamuy/etal:2002})
and IAU circulars. This difficulty is inherent to the SN
classification scheme, rather than from a physical mis-conception of
these events, both of which are believed to originate in the core
collapse of a massive star, stripped of its outer layers
through either stellar winds or interaction with a binary companion 
\citep{Woosley/Langer/Weaver:1993,Woosley/Langer/Weaver:1995}. SNe~Ib
are defined by the presence of conspicuous lines of \heone\ in their 
optical spectra, whereas SNe~Ic are defined by their
quasi-absence \citep{Matheson/etal:2001}. The \sitwo\
\l6355 feature is weak in SNe~Ib/c, which enables one to
differentiate them from SNe~Ia, at least in principle (see
\S~\ref{Sect:1cnorm1a}). Both are of Type I and are thus also defined
by the absence of hydrogen lines in their spectra, although the case
has recently been made for some hydrogen being present in SNe~Ib/c 
\citep{Branch/etal:2006,Elmhamdi/etal:2006}.

\begin{figure}
\epsscale{1.15}
\plotone{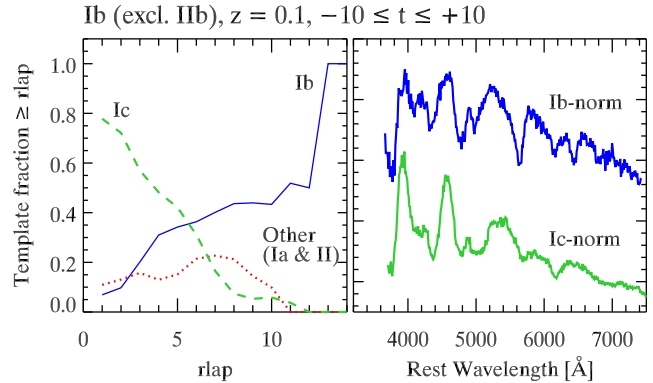}
\caption{Attempt to classify an SN~Ib at $z=0.1$ in the 
  age interval $-10 \le t_V \le +10$. The lines
  correspond to fractions of matching templates of Type Ib ({\it solid
  line}), Type Ic ({\it dashed line}) and all other types ({\it
  dotted line}), in $r{\rm lap}$ bins of size unity. The Type IIb subclass,
  included in both the SN~Ib and SN~II categories, is excluded for
  this purpose. The right panel shows representative spectra of
  normal SN~Ib and Ic around maximum light. Note that
  the relative differences in {\it pseudo}-continuum shapes have no
  impact on the \snid\ results. {\it [See the electronic version of
  the Journal for a color version of this figure.]} 
\label{Fig:Ib}}
\end{figure}

In Fig.~\ref{Fig:Ib}, we show the result of cross-correlating SN~Ib
 spectra at low redshift ($z=0.1$) within 10 days from
$V$-band maximum with SNe of all types in the \snid\ database:
Type Ib (excluding Type IIb; {\it solid line}), Type Ic ({\it dashed
  line}), and other SN types ({\it dotted line}). The lines correspond to
the fraction of template supernovae of a given type, as a 
function of the $r{\rm lap}$ quality parameter. We deliberately exclude the IIb
subtype from this analysis, as this type is a hybrid between the
 II and Ib subtypes (see below). The results of Fig.~\ref{Fig:Ib}
are encouraging: for $r{\rm lap} > 6$, the recovered fraction of
SN~Ib dominates over SNe~Ic, with less than 25\%\
confusion with other supernova types.
For $r{\rm lap} \ge 13$, only Type Ib templates correlate with
the input spectrum.

\subsubsection{SN~IIb versus SN~II/Ib}

Some supernovae ``evolve'' from one type to another as they age. 
Such is the case for SNe~IIb 
\citep{Woosley/etal:1987}, whose early-time spectrum is characterized
by prominent P Cygni lines of the hydrogen Balmer series (as in
SNe~II), but which later develop conspicuous lines of
\heone, as in SNe~Ib. A proto-typical example of such a
supernova is SN~1993J \citep{Nomoto/etal:1993,Matheson/etal:2000}. The
\snid\ database currently contains spectra for three SNe~IIb:
SN~1993J, SN~2000H and SN~2006T (Table~\ref{Table:sndb}).

\begin{figure}
\epsscale{1.15}
\plotone{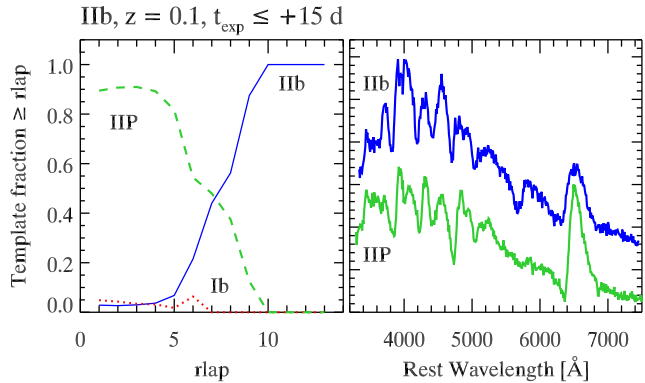}
\caption{Same as Fig.~\ref{Fig:Ib}, but for an SN~IIb at
  $z=0.1$ in the age interval $t_{\rm exp} \le +15$
  ($t_{\rm exp}$ is the number of days past explosion). The lines
  correspond to fractions of matching templates of Type IIb ({\it solid
  line}), Type II ({\it dashed line}) and Type Ib ({\it dotted
  line}), in $r{\rm lap}$ bins of size unity. {\it [See the electronic
  version of the Journal for a color version of this figure.]}
\label{Fig:IIb}}
\end{figure}

In Fig.~\ref{Fig:IIb} we study the fraction of templates corresponding
to Type IIb ({\it solid line}), Type II ({\it dashed line}), and Type Ib
({\it dotted line}), when the input spectrum is a Type IIb at low
redshift ($z=0.1$), within 15 days past explosion. At these ages,
the \heone\ lines are somewhat weaker than after maximum and the
confusion with SNe~II is greater: only for $r{\rm lap} \gtrsim
7$ is the fraction of recovered IIb templates greater than ordinary
SNe~II. The confusion with Type Ib templates is small ($< 20\%$)
up to $r{\rm lap} \approx 9$ and null for larger values of
$r{\rm lap}$. Templates corresponding to SNe~Ia and Ic do not correlate
well with input Type IIb spectra and the confusion is
practically nonexistent ($< 5\%$, not shown here). Again, more SNe~IIb
are needed to truly test the ability for \snid\ to
correctly identify them.

\subsection{Using \snid\ for SN Identification}

The previous examples illustrate the ability of \snid\ to recover a
significant fraction of supernovae in the database corresponding to
the input SN type. While Figs.~\ref{Fig:Ia91t}--\ref{Fig:IIb} are
informative, they do not provide a unique answer to the following: how
does one use \snid\ to determine the type of an SN spectrum and
can one relate the $r{\rm lap}$ quality parameter to a formal confidence that an
input spectrum is of a certain type (and, sometimes more importantly,
that it is not of another type)? The answer is far from settled,
and its resolution will probably involve a more sophisticated Bayesian
approach (see next section). Nevertheless, we have tested the following
classification schemes:

\begin{itemize}

\item[1.]{The SN type (subtype) is simply determined as the type
  (subtype) of the best-match template(s) for $r{\rm lap}$ greater than some
  cutoff value $r{\rm lap}_{\rm min}$.}

\item[2.]{The SN type (subtype) is the one corresponding to the
  highest fraction of templates corresponding to a given type
  (subtype) for $r{\rm lap} \ge r{\rm lap}_{\rm min}$, with the possible
  additional requirement that this fraction exceeds some cutoff.}

\item[3.]{The SN type (subtype) is the one whose fraction increases
  most with increasing $r{\rm lap}$ (i.e. the lines shown in
  Figs.~\ref{Fig:Ia91t}--\ref{Fig:IIb} that have the highest positive
  gradient for $r{\rm lap} \ge r{\rm lap}_{\rm min}$), with the possible
  additional requirement that this gradient exceeds some cutoff.}

\end{itemize}

The first of these is the one commonly used for the identification of
supernovae, at both low and high redshift. In IAU circulars, a
supernova is announced to be of a given type when it is ``(most)
similar to supernova $X$ at $N$ days from maximum light.'' In high-$z$
SN~Ia searches, a secure type is determined when a given spectrum is
sufficiently similar to a nearby SN~Ia (e.g.,
\citealt{Matheson/etal:2005}), but it is never clear how {\it 
different} it is from supernovae of other types. However, the
best-match template is not always the best indicator of the SN type
(see the distinction between post-maximum spectra of 1991T-like SNe~Ia
and other SNe~Ia in Fig.~\ref{Fig:Ia91t}) and the large spectral
database used in \snid\ offers the possibility to use the statistical
power of correlations exceeding a certain $r{\rm lap}$ cutoff. This second
classification scheme has already been used for the classification of
high-$z$ SN~Ia spectra using \snid\ \citep{Miknaitis/etal:2007}. A
drawback of such an approach is that the determination of the SN type
depends on the completeness of the SN database--- which comprises few 
template spectra of core-collapse SNe (Ib,Ic, and II; see
\S~\ref{Sect:sniddb}). Thus, for instance, it is not possible to
identify peculiar SNe~Ia this way (Fig.~\ref{Fig:Iapec}). Last, the
``gradient method'' for classification is generally more robust, but
also requires some other constraint on either the template fraction or
the type of the best-match template(s).

We are extensively testing these different combinations, to
optimize the type determination for all SN types, at varying $z$, $t$,
and S/N, although we suspect that a more elaborate Bayesian treatment will
be required to properly account for the probability of an input
spectrum to be of a certain type (as well as {\it not} being of some
other type).


\section{Comparison with Other Methods and Further Improvements}\label{Sect:comparison}

Several other methods are used to determine the type,
redshift, or age of a supernova spectrum. We briefly describe them
here, with a distinction between cross-correlation and
``$\chi^2$-based'' methods. Last, we comment on the alternative
Bayesian approach to supernova classification--- only applied so far
to photometric measurements and its possible implementation in \snid.

Other spectral classification methods involve principle component
analysis (PCA), possibly in combination with artificial neural
networks. These are beyond the scope of this paper and we do not
discuss them here, although the PCA method has already been applied to SN~Ia
classification by \citet{James/etal:2006}.

\subsection{Cross-correlation Methods}

We are aware of two other algorithms based on the cross-correlation
techniques of \citet{Tonry/Davis:1979}. Both are aimed at determining
redshifts of galaxies (or stars) in large surveys, but could easily
be tuned (e.g., by modifying the shape of the bandpass
filter and including age information) to supernovae as in \snid.

XCSAO \citep{Kurtz/Mink:1998} is a program part of the
IRAF\footnote{IRAF is distributed by the National Optical Astronomy
Observatories, operated by the Association of Universities for
Research in Astronomy, Inc., under contract to the National Science
Foundation of the United States.} RVSAO package, aimed at obtaining 
redshifts and radial velocities from digital spectra. It has been used
extensively in the past for redshift surveys (for references see
\citealt{Kurtz/Mink:1998}) and is currently used in
the Smithsonian Hectospec Lensing Survey (SHELS;
\citealt{Geller/etal:2005}). The basic algorithm is the same as that
described in \S~\ref{Sect:xcorform}, although some important
differences exist. First, the overlap in wavelength between the input and
template spectra at the correlation redshift (the ${\rm lap}$ parameter
discussed in \S~\ref{Sect:lap}) is not taken into account
explicitely; rather it is maximized by including spectral templates
at different redshifts. Second,
XCSAO directly selects the best peak in the correlation
function, rather than looking at the 10 best peaks individually. Last,
the reported correlation redshift is that associated with the
best-match template (i.e., the one with the highest correlation
  height-noise ratio, $r$), rather
than being the median of all redshifts above a certain cutoff
$r_{\rm min}$. It may be that not including the ${\rm lap}$ parameter is less 
important for galaxy redshift determination, due to the narrower
widths of spectral lines in galaxy spectra and the iterative scheme
implemented in XCSAO ensures that an optimal correlation redshift is
computed (D.~Mink 2007, private communication). For supernova spectra, however, the
inclusion of the ${\rm lap}$ parameter is fundamental to obtain reliable
redshifts (see \S~\ref{Sect:zreserr}). The other two differences are
of a less fundamental nature.

Another algorithm based on the correlation techniques of
\citet{Tonry/Davis:1979} is RUNZ
\citep{Saunders/Cannon/Sutherland:2004}, a program used by members of
the 2dF Galaxy Redshift Survey \citep{Colless/etal:2001} and the 6dF
Galaxy Survey \citep{Jones/etal:2004}. An essential difference with
\snid\ is the scaling of the input spectrum by its inverse-variance
(see \S~\ref{Sect:xcorz}), which leads to improved
cross-correlation redshifts for galaxy spectra. As mentioned
earlier in \S~\ref{Sect:xcorz}, no such improvement is found for
supernova redshifts, since the power spectrum of a typical variance
spectrum (for ground-based observations) peaks at higher wavenumbers
than for SN spectra. One advantage of RUNZ over \snid\ is the
implementation of Gaussian constraints on redshift, a feature that will
be part of a future version of \snid.

\subsection{$\chi^2$ Methods}

An alternative to cross-correlation techniques involves the
minimization of a $\chi^2$-like quantity at discrete redshift
intervals, to find the best match between an input and template
spectrum. Such techniques do not enable a formal evaluation of the
redshift error as in \snid\ (see \S~\ref{Sect:zerr}). An elaborate
implementation of this approach is ``superfit''
\citep{Howell/etal:2005}, in which an input spectrum is compared to a
combination of a (possibly reddened) template supernova spectrum and a
template galaxy spectrum at a given redshift. A similar program to
superfit is $\mathcal{SN}$-fit (\citealt{Sainton:2004}; see also
discussion in \citealt{Blondin/etal:2005}). The number of free
parameters means that the execution time is 2--3 orders of magnitude greater
than for \snid. For this reason we could not test superfit in the
same way we conducted the simulations presented in the previous
sections.

The advantage of this technique lies in the evaluation of the
fraction of galaxy light in the input spectrum and its
subsequent subtraction to obtain a ``pure'' supernova
spectrum. 
Consequently, supernovae can be classified even when the galaxy
contamination fraction is high ($\lesssim 75\%$;
\citealt{Howell/etal:2005}).
Given the impact of galaxy contamination on 
correlation redshifts (Fig.~\ref{Fig:zvar_gf}), we expect more
accurate redshifts when this contamination is removed 
from the input spectrum. However, the typical errors on redshift are
similar to those in \snid\ 
\citep{Hook/etal:2005}. This suggests that the 
redshift accuracy is limited by physical properties of supernovae
(namely, the velocity location of their prominent spectral features)
rather than by differences in algorithm. The accuracy on the
age determination is similar to that in \snid, with a $\sim 3$-day
dispersion about the one-to-one correspondence with the light-curve
age \citep{Hook/etal:2005,Howell/etal:2005}. We note that the
current version of superfit has not been extensively tested for
redshift and age determination and has not been tested at all for
type determination (A.~Howell 2006, private communication).

A variant of the $\chi^2$ approach implemented in superfit is the
spectral feature aging (SFA) algorithm of \citet{Riess/etal:1997},
used to determine the age of a normal SN~Ia of known
redshift. In SFA, the input spectrum is divided up into several
wavelength intervals (typically eight) and each of these spectral
regions is compared with corresponding ones in a database of template
spectra. The age accuracy is similar to that in \snid but is
largely sensitive to the wavelength intervals of the spectral regions
used to divide up the input spectrum \citep{Foley/etal:2005}. Another
tool, based on the SFA algorithm, plans to extend the age
determination to spectra of all types \citep{Harutyunyan/etal:2005}.

\subsection{Bayesian Approach to SN Classification}

Recently, several authors have presented Bayesian methods to
determine the type of a supernova based on a single-epoch photometric
measurement (potentially in multiple bands;
\citealt{Poznanski/Maoz/Gal-Yam:2006}), or on multiband light curves
\citep{Kuznetsova/Connolly:2007,Kunz/etal:2006}. The motivation behind
these purely photometry-based approaches is the planned
next-generation of wide-field all-sky surveys (such as Pan-STARRS and
LSST), for which many SN~Ia candidates will be too faint for
spectroscopy. However, these techniques are in principle   
applicable to ongoing high-redshift SN~Ia surveys, which are limited
by the amount of available spectroscopy time
\citep{Matheson/etal:2005,Howell/etal:2005}. In fact,
\citet{Tonry/etal:2003} already adopt a Bayesian approach in
fitting high-$z$ SN~Ia light curves with BATM (Bayesian Adapted
Template Match; \citealt{Tonry/etal:2003}).

These Bayesian-based approaches assign a probability for a supernova
to be of a certain type, based on a set of measurements (e.g., light
curve points in a given photometric band) and given a model--- or
template, that depends on a set of parameters. As pointed out by
\citet{Kuznetsova/Connolly:2007}, when computing this probability it
is generally assumed that the input is indeed a supernova of a known
type, although in principle one could extend the formalism to
incorporate all known astrophysical objects. Moreover, such methods
invoke ``marginalization over type,'' which poses some conceptual
problems, since it assumes the SN classification scheme to be both
complete (i.e. to include all possible SN types) and 
hermetic. Yet, there appears to be a continuum of properties relating
different SN types (e.g., Type Ib and Ic), some supernovae evolve from
one type to another (e.g., Type IIb) and others still seem to defy
classification (such as the ``peculiar SNe~Ia'' SN~2002cx or SN~2005hk).

The current version of \snid\ does not incorporate such a methodology in
its supernova classification. Nevertheless, the change in the relative
fraction of templates of a given type as a function of the $r{\rm lap}$
quality parameter could be folded in as an extra constraint on the SN
type in a Bayesian framework.


\section{Conclusion}\label{Sect:ccl}

We have presented an algorithm, based on the correlation techniques of
\citet{Tonry/Davis:1979}, that can be used to
determine the redshift and age of a supernova spectrum and place
constraints on its type. We develop a diagnostic, the $r{\rm lap}$
quality parameter, to quantify the reliability of a given correlation between the
input and a template spectrum. This parameter is simply the
product of the \citet{Tonry/Davis:1979} correlation height-noise ratio ($r$) and the overlap in
rest-frame $\ln \l$ space between the input and template spectrum at the 
correlation redshift (${\rm lap}$). This $r{\rm lap}$ diagnostic has the advantage of
enabling the formal computation of the redshift error, proportional to
$1/(1+r{\rm lap})$. We show, based on simulations, that for $r{\rm lap} 
\gtrsim 5$, the typical error on redshift and age is $\sigma_z \lesssim
0.01$ and $\sigma_t \lesssim 3$ days, respectively. The former
accuracy on redshift is confirmed through a comparison of correlation
redshifts with host-galaxy redshifts (determined from narrow lines in
the spectrum) out to redshifts $z \lesssim 0.8$. The latter accuracy
on age is confirmed through a comparison of (rest frame) spectral
ages using \snid\ and (observer frame) light curve ages corrected
for the $(1+z)$ time-dilation factor expected in an expanding
universe. The fact that both age estimates agree so well is itself a
verification of the cosmological nature of redshifts, previously
tested with multiepoch SN~Ia spectra
\citep{Riess/etal:1997,Foley/etal:2005}. Furthermore, the success of
\snid\ in determining the redshift and age of the high-redshift SN~Ia
spectra in our sample shows that these are similar to
  low-redshift counterparts.

We present first results of an impartial and effective spectroscopic
classification of supernovae, based on the fraction of correlations
exceeding a certain $r{\rm lap}$ cutoff. We illustrate this through various
examples, three of which are relevant to ongoing SN~Ia searches at
high redshift: we are able to distinguish 1991T-like SNe~Ia from other
SNe~Ia at $z=0.5$;
we identify SNe~Ib/c as such at both $z=0.3$ and
$z=0.5$. The identification of peculiar SNe~Ia, on the other
hand, proves easier at $z=0.5$ than $z=0.3$, although this
result remains to be verified with more peculiar SN~Ia spectral
templates. In all cases we assume the galaxy contamination fraction,
to which \snid\ is {\it not} sensitive, to be negligible.
 
These examples both illustrate the success and 
limitations of such an automated classification scheme and highlight
the complementarity between spectroscopic and photometric observations
in determining the supernova type. We are currently testing various
combinations of other classification schemes to improve the
classification of all SN types using \snid.


Supernova discoveries will continue to increase dramatically with the
advent of wide-field imaging telescopes optimized for the detection of
transient events, such as Pan-STARRS and LSST. These experiments
are expected to find tens to hundreds of thousands of new supernovae
each year, few of which will have spectroscopic
confirmation. Thus, most identifications will have to rely solely on
photometric properties, a difficult task in view of the present
difficulty of distinguishing between SN types (and subtypes) with
spectra. It is likely that those experiments will have to rely on a
simpler classification scheme, focusing on the main SN types (Type Ia,
Ib/c, \& II) with little or no distinction between the associated
subtypes.

Future planned space-based high-redshift SN~Ia surveys within the
NASA/DOE Joint Dark Energy Mission (ADEPT, DESTINY, SNAP) will
incorporate a spectrograph and could benefit from a tool such as
\snid. A secure identification of SNe~Ia in such experiments will
require sufficient rest frame wavelength coverage beyond $\sim 5500$\,\AA,
as the distinction between SNe~Ia and Ic (and between SN~Ia subtypes,
including peculiar events) is otherwise problematic
(Figs.~\ref{Fig:Ia91t}--\ref{Fig:Iapec}).

The current version of \snid\ will be made available to the
community and we will set up a Web-based interface for
instantaneous supernova typing (including redshift and age
determination). Future versions of \snid\ will include a
wavelength- and age-weighted spectrum overlap parameter, ${\rm lap}(\l,t)$, an explicit 
treatment of the covariance between redshift and age and a Bayesian
approach to type determination, as currently used for photometric
classification of supernovae
\citep{Poznanski/Maoz/Gal-Yam:2006,Kuznetsova/Connolly:2007}.
Moreover, more spectral templates are continuously being included in
the \snid\ database through the CfA Supernova Program (more than 3000
spectra of over 700 supernovae since 1997), which directly
impact the ability of \snid\ to securely identify input spectra. This
further enables comparative studies of SN spectra and quantitative
evaluation of synthetic spectral fits to observations.


\acknowledgments
The authors wish to thank Tom Matheson for use of CfA SN~Ia
spectra ahead of publication, as well as the ESSENCE team for providing
spectra and light curves of the high-redshift SNe~Ia used in
Fig.~\ref{Fig:zgaltlcsnid}. We also thank Weidong Li and Rubina Kotak
for providing spectra of SN~2002cx and SN~2002er, respectively. The
present paper benefited from discussions with the CfA Supernova Group,
and we thank Malcolm Hicken for his work on light-curve data that led
to the inclusion of several extra supernovae in the \snid\
database. We thank the referee andy Howell, for his comments
that led to an improvement of the paper and for making available to
us the spectrum of SNLS-03D3bb \citep{Howell/etal:2006}.
We also thank Robert Kirshner, Bruno Leibundgut, Brian Schmidt and
Kevin Krisciunas for useful comments on the manuscript.

S.~B. thanks the Kavli Institute for Theoretical Physics at
the University of California, Santa Barbara for its hospitality during
the program: ``Accretion and Explosion: the Astrophysics of
Degenerate Stars.''

This work has been funded in part by the US National Science
Foundation through grants AST 04-43378, AST 05-7475, AST
06-06772 and PHY 05-51164. This research has made use of the CfA
Supernova Archive, which is funded in part by the National Science
Foundation through grant AST 06-06772.


\clearpage
\LongTables
\begin{deluxetable*}{llll}
\tabletypesize{\scriptsize}
\tablenum{1}
\tablewidth{0pt}
\tablecaption{\snid\ Supernova Database\label{Table:sndb}}
\tablehead{
\colhead{IAU Name} & 
\colhead{Subtype} & 
\colhead{Ages}&
\colhead{Ref.} \\
\colhead{(1)} & 
\colhead{(2)} & 
\colhead{(3)} & 
\colhead{(4)}
}
\startdata
1981B  & Ia-norm  & 0,17,20,24                                                                                                        & 1              \\
1986G  & Ia-91bg  & $-$3,$-$2,0--2,30+(4)                                                                                             & 1              \\ 
1987A  & II-pec   & 2,4--9,11--27,31--39,40+(69)                                                                                      & 2,3            \\ 
1989B  & Ia-norm  & $-$6,$-$1,4,6,8,10,12(2),13,14,16--25,30+(3)                                                                      & 1              \\ 
1990B  & Ic-norm  & 5,6(2),7,9,10,15,28(2),30+(6)                                                                                     & 4,5,CfA        \\ 
1990I  & Ib-norm  & 11,19,30+(6)                                                                                                      & 6              \\ 
1990N  & Ia-norm  & $-$13(2),$-$6,3,5,15,18,30+(5)                                                                                    & 7,8            \\ 
1990O  & Ia-norm  & $-$[7--5],0,19,20                                                                                                 & CfA            \\
1991M  & Ia-norm  & 25,26,30+(1)                                                                                                      & CfA            \\ 
1991T  & Ia-91T   & $-$12,$-$10,$-$9,$-$[7--5],0,19,30+(3)                                                                            & 9--11          \\ 
1991bg & Ia-91bg  & 2,3(3),15,16,19(2),20,26,30+(16)                                                                                  & 12,14          \\ 
1992A  & Ia-norm  & $-$5(2),$-$1,0,2,3,6(2),7,9(2),12(2),16,17,24,28                                                                  & 15             \\
1992H{\tablenotemark{a}} & IIP & 16,29,40+(8)                                                                                         & 16,17          \\ 
1992ar & Ic-norm  & 3                                                                                                                 & CfA            \\
1993J  & IIb      & 3,4(2),5,11,16,17,18,19(3),22,24(2),25(2),26--28,32--34,38,40+(51)                                                & 18--21         \\ 
1993ac & Ia-norm  & 7                                                                                                                 & CfA            \\
1994D  & Ia-norm  & $-$11(2),$-$10(2),$-$9,$-$8,$-$6,$-$5(2),$-$4(2),$-$3,$-$2,0,2,3,10--12,13(2),14,15(3),16,17(2),19,24,26,30+(11)  & 22,23,CfA      \\ 
1994I  & Ic-norm  & $-$6(2),$-$4,$-$3,1,2(2),3,21--24,26,30+(5)                                                                       & 24             \\ 
1994M  & Ia-norm  & 3--5,8,13,14                                                                                                      & CfA            \\
1994Q  & Ia-norm  & 19                                                                                                                & CfA            \\
1994S  & Ia-norm  & $-$3(2),1                                                                                                         & CfA            \\
1994T  & Ia-norm  & 1                                                                                                                 & CfA            \\
1994ae & Ia-norm  & 1,2,3(2),4,6,9(2)10,11,30+(7)                                                                                     & CfA            \\ 
1995D  & Ia-norm  & 4,6,8,10,11,14,16,30+(3)                                                                                          & CfA            \\ 
1995E  & Ia-norm  & $-$2,0,2,7,10,30+(1)                                                                                              & CfA            \\ 
1995ac & Ia-norm  & 24                                                                                                                & CfA            \\
1995al & Ia-norm  & 17,25                                                                                                             & CfA            \\
1995bd & Ia-norm  & 11,21,30+(2)                                                                                                      & CfA            \\ 
1996C  & Ia-norm  & 8                                                                                                                 & CfA            \\
1996X  & Ia-norm  & $-$3,$-$2,$-$1(2),0,1(2),2(2),3,5--7,8(2),9,13,21,23,25,30+(1)                                                    & 25,CfA         \\ 
1997br & Ia-91T   & $-$9,$-$8,$-$7(2),$-$6(2),$-$4,8,9,12,17,18,21,24,30+(6)                                                          & 26,CfA         \\ 
1997cn & Ia-91bg  & 4,29,30+(1)                                                                                                       & 27,CfA         \\ 
1997do & Ia-norm  & $-$11,$-$10,$-$7,$-$6,9,11,12,13,15,16,20,21                                                                      & 28             \\
1997dt & Ia-norm  & $-$[10--7],$-$4,1,3                                                                                               & 28             \\
1997ef & Ic-hyper & $-$14,$-$12,$-$[11--9],$-$6,$-$5(2),$-$4,7,13,14,16,17,19,22,24,27,30+(4)                                         & 29             \\ 
1998S  & IIn      & 5,6,17,19,20(2),21,28,30--32,34,40+(44)                                                                           & 30--32         \\ 
1998V  & Ia-norm  & 1--3,13,14,15,30+(3)                                                                                              & 28             \\ 
1998ab & Ia-91T   & $-$7,7,8,18,19,20,21,22,23,30+(3)                                                                                 & 28             \\ 
1998aq & Ia-norm  & $-$9,$-$8,0--7,19,21,30+(15)                                                                                      & 28,33          \\ 
1998bp & Ia-91bg  & $-$2,$-$1,0--2,13,15,25,26,28,30+(1)                                                                              & 28             \\ 
1998bu & Ia-norm  & $-$[3--1],1,9--14,28,29,30+(21)                                                                                   & 28,34          \\ 
1998bw & Ic-hyper & 8,9,12--14,16,18,19,21,24,26--28,30+(8)                                                                           & 35             \\ 
1998de & Ia-91bg  & $-$[7--5],$-$3,$-$2,0,3                                                                                           & 28             \\
1998dh & Ia-norm  & $-$[9--7],$-$5,$-$3,0,30+(4)                                                                                      & 28             \\ 
1998dk & Ia-norm  & 10,11,13,16,18,21,23,30+(2)                                                                                       & 28             \\ 
1998dm & Ia-norm  & 5,6,8,11,13,16,18,25,30+(2)                                                                                       & 28             \\ 
1998dt & Ib-norm  & 0,1,4,7,11,12,17                                                                                                  & CfA            \\
1998ec & Ia-norm  & $-$2,$-$1,13,21,27,30+(1)                                                                                         & 28             \\ 
1998eg & Ia-norm  & 0,5,18,20,23                                                                                                      & 28             \\
1998es & Ia-91T   & $-$[10--1],1--3,16,18,19,20,24,26,30+(7)                                                                          & 28             \\ 
1999X  & Ia-norm  & 12,13,15,16,21,29                                                                                                 & 28             \\
1999aa & Ia-91T   & $-$[9--1],1,15--18,27--29,30+(11)                                                                                 & 28             \\ 
1999ac & Ia-91T   & $-$4,$-$3,$-$1,9--12,25,27,29,30+(9)                                                                              & 28             \\ 
1999by & Ia-91bg  & $-$[5--2],2--8,11,25,29,30+(3)                                                                                    & 28,36          \\ 
1999cc & Ia-norm  & $-$3,$-$1,0,2,19,24,25                                                                                            & 28             \\
1999cl & Ia-norm  & $-$8,$-$7,$-$[5--1],1,8,30+(1)                                                                                    & 28             \\ 
1999dq & Ia-91T   & $-$[10--2],1--4,6,18,19,27,30+(6)                                                                                 & 28             \\ 
1999ee & Ia-norm  & $-$9,$-$7,0,2,7,11,16,20,22,27,30+(2)                                                                             & 28             \\ 
1999ej & Ia-norm  & $-$1,2,4,9,12                                                                                                     & 28             \\
1999em & IIP      & 6(2),7--9,10(2),11,12,15,16(3),17,19,21,26,37,40+(27)                                                             & 37--39         \\ 
1999ex{\tablenotemark{b}} & Ib-norm  & $-$5,0,9                                                                                       & 40             \\
1999gd & Ia-norm  & 3,9,27,30+(2)                                                                                                     & 28             \\ 
1999gh & Ia-91bg  & 5--9,11,12,28,30+(7)                                                                                              & 28             \\ 
1999gi & IIP      & 5,7,8,31,36,39,40+(5)                                                                                             & 41             \\ 
1999gp & Ia-91T   & $-$5,$-$2,0,3,5,7,22,30+(3)                                                                                       & 28             \\ 
2000B  & Ia-norm  & 9,14,22,30+(2)                                                                                                    & 28             \\ 
2000E  & Ia-norm  & $-$6,$-$3,$-$1,8,30+(1)                                                                                           & 42             \\ 
2000H  & IIb      & 28,29(2),31--34,40+(5)                                                                                            & 43,CfA         \\ 
2000cf & Ia-norm  & 3,4,15,17,25,26                                                                                                   & 28             \\
2000cn & Ia-91bg  & $-$[9--7],9,11,13,22,26,27,30+(1)                                                                                 & 28             \\ 
2000cx & Ia-pec   & $-$[3--1],0--2,6--8,10,12,15,19,22,24,26,28,30+(9)                                                                & 44             \\ 
2000dk & Ia-91bg  & $-$5,$-$4,1,4,10,30+(3)                                                                                           & 28             \\ 
2000fa & Ia-norm  & $-$10,$-$9,2,3,5,9,11,14,16,18,21,30+(3)                                                                          & 28             \\ 
2001V  & Ia-norm  & $-$[13--9],$-$[7--5],$-$3,10,11(2),13,14,18,19,20(2),21(2),22--24,27,28,30+(13)                                   & 28             \\ 
2002ap & Ic-hyper & $-$6,$-$5,$-$2,$-$1,0--2,4--6,7(2),10,12,26,30+(5)                                                                & 45,CfA         \\ 
2002bo & Ia-norm  & $-$13(2),$-$12(2),$-$[9--6],$-$5(2),$-$4(2),$-$3(3),$-$2(2),$-$1(3),0,6,11--22,24,28,29(2),30+(12)                & 46,CfA         \\ 
2002cx & Ia-pec   & $-$5,$-$2,10,14,18,23,24,30+(1)                                                                                   & 47             \\ 
2002er & Ia-norm  & $-$11,$-$[9--5],$-$4(2),$-$[3--1],2--4,6,9,11,12,16(2),20,24,30+(2)                                               & 48             \\ 
2003du & Ia-norm  & $-$12,$-$10(3),$-$[9--5],$-$3,$-$1(2),0--2,3(2),4,5,8--11,14,17--19,20(2),22(2),24,26,27,29,30+(24)               & 49,50          \\ 
2004ao & Ib-norm  & 7--13,16,17,20--23,30+(16)                                                                                        & CfA            \\ 
2004aw & Ic-norm  & $-$5,$-$3,$-$2(2)$-$1,0,2(2),3--5,12,18(2),19,21,23,25,26,30+(6)                                                  & 51,CfA         \\ 
2004et & IIP      & 13--15,17,18,20,22,26,28,40+(15)                                                                                  & CfA            \\ 
2004gt & Ic-norm  & 15,17,21,23,30+(8)                                                                                                & CfA            \\ 
2005bf{\tablenotemark{c}} & Ib-pec   & $-$4,$-$2,$-$1,0,2,16,18--23,25--27,29,30+(6)                                                  & 52             \\ 
2005cs & IIP      & 7--14,16--19,35,36,40+(1)                                                                                         & 53,54          \\ 
2005hg & Ib-norm  & $-$[13--1],0,13,17,27                                                                                             & CfA            \\
2005hk & Ia-pec   & $-$9$-$8(2),$-$7(2),$-$6(3),$-$5(2),$-$4(3),$-$3(2),$-$2,$-$1,4,12,14(2),20,23(2),26,27,30+(6)                    & 55,CfA         \\ 
2005mf & Ic-norm  & $-$4,$-$3,3,6                                                                                                     & CfA            \\
2005mz & Ia-91bg  & $-$7,11,13,19                                                                                                     & CfA            \\
2006T  & IIb      & 8,10,28,36                                                                                                        & CfA            \\
2006aj & Ic-broad & $-$5,$-$3,$-$2,$-$1,0,2,3                                                                                         & 56             \\
\enddata
\tablenotetext{\ }{
{\it Column headings:}
(1) IAU designation.
(2) Supernova subtype, as defined in Table~\ref{Table:stype}.
(3) Rest-frame SN age, rounded to closest whole day, in days from
$B$-band maximum (for SN~Ia), from $V$-band maximum (for SN~Ib/c), or from the 
estimated date of explosion (for SN~II). Adjacent ages are listed in
between square brackets; a ``(n)'' indicates that $n$ spectra
correspond to a same rounded age. Spectra of SN~Ia/b/c whose age
exceeds +30 days are grouped together, e.g. 30+(5) indicated there are
5 spectra with ages $\ge +30$ days (past maximum); same for spectra of SN~II whose
age exceeds 40 days (past explosion).
(4) Reference of refereed articles presenting optical spectroscopic
data (see ``References'' below); ``CfA'' refers to unpublished spectra
obtained by members of the CfA SN Group, some of which are available
{\it via} the CfA Supernova Archive ({\tt
http://www.cfa.harvard.edu/supernova/SNarchive.html}).
}
\tablenotetext{a}{The light curve of SN~1992H exhibited a truncated
plateau \citep{Clocchiatti/etal:1996}, but its spectra are otherwise
indistinguishable from other Type IIP supernovae.}
\tablenotetext{b}{\citet{Hamuy/etal:2002} classify SN~1999ex as an
  intermediate Ib/c event, while \citet{Parrent/etal:2007} support the
  Ib classification, highlighting the similarity with the peculiar
  SN~Ib 2005bf \citep{Tominaga/etal:2005}. We classify SN~1999ex as a
  normal Type Ib supernova and note that the essential spectroscopic
  peculiarity of SN~2005bf (namely the increasing absorption velocity
  of the \heone\,\l5876 line; \citealt{Tominaga/etal:2005}) is not
  present in SN~1999ex.}
\tablenotetext{c}{The $V$-band light curve of SN~2005bf had two
distinct maxima. The age is expressed in days from the {\it first}
maximum.}
\tablerefs{
(1)  \citealt{Wells/etal:1994}; 
(2)  \citealt{Phillips/etal:1988};
(3)  \citealt{Phillips/etal:1990}; 
(4)  \citealt{Matheson/etal:2001};
(5)  \citealt{Clocchiatti/etal:2001};
(6)  \citealt{Elmhamdi/etal:2004};
(7)  \citealt{Leibundgut/etal:1991};
(8)  \citealt{Mazzali/etal:1993};
(9)  \citealt{Jeffery/etal:1992};
(10) \citealt{Schmidt/etal:1994};
(11) \citealt{Mazzali/etal:1995};
(12) \citealt{Leibundgut/etal:1993};
(13) \citealt{Turatto/etal:1996};
(14) \citealt{Gomez/Lopez/Sanchez:1996};
(15) \citealt{Kirshner/etal:1993};
(16) \citealt{Clocchiatti/etal:1996};
(17) \citealt{Gomez/Lopez:2000};
(18) \citealt{Jeffery/etal:1994};
(19) \citealt{Barbon/etal:1995};
(20) \citealt{Matheson/etal:2000};
(21) \citealt{Fransson/etal:2005};
(22) \citealt{Hoeflich:1995};
(23) \citealt{Patat/etal:1996};
(24) \citealt{Millard/etal:1999};
(25) \citealt{Salvo/etal:2001};
(26) \citealt{Li/etal:1999};
(27) \citealt{Turatto/etal:1998};
(28) \citealt{Matheson/etal:2007}; 
(29) \citealt{Iwamoto/etal:2000};
(30) \citealt{Lentz/etal:2001};
(31) \citealt{Fassia/etal:2001};
(32) \citealt{Fransson/etal:2005};
(33) \citealt{Branch/etal:2003};
(34) \citealt{Jha/etal:1999};
(35) \citealt{Patat/etal:2001};
(36) \citealt{Garnavich/etal:2004};
(37) \citealt{Baron/etal:2000};
(38) \citealt{Hamuy/etal:2001};
(39) \citealt{Leonard/etal:2002a};
(40) \citealt{Hamuy/etal:2002};
(41) \citealt{Leonard/etal:2002b};
(42) \citealt{Valentini/etal:2003};
(43) \citealt{Branch/etal:2002};
(44) \citealt{Li/etal:2001a};
(45) \citealt{Gal-Yam/Ofek/Shemmer:2002};
(46) \citealt{Benetti/etal:2004a};
(47) \citealt{Li/etal:2003};
(48) \citealt{Kotak/etal:2005};
(49) \citealt{Anupama/Sahu/Jose:2005};
(50) \citealt{Stanishev/etal:2007};
(51) \citealt{Taubenberger/etal:2006};
(52) \citealt{Tominaga/etal:2005};
(53) \citealt{Brown/etal:2007};
(54) \citealt{Dessart/etal:2007};
(55) \citealt{Phillips/etal:2007};
(56) \citealt{Modjaz/etal:2006}
}
\end{deluxetable*}
\clearpage

\end{document}